\documentclass[%
 reprint,
superscriptaddress,
 amsmath,amssymb,
 aps,
prb,
floatfix,
]{revtex4-2}

\makeatletter
\newcommand{\fmarki}{*}
\newcommand{\fmarkii}{\ensuremath{\dagger}}
\newcommand{\fmarkiii}{\ensuremath{\ddagger}}
\newcommand{\fmarkiv}{\ensuremath{\mathsection}}
\newcommand{\fmarkv}{\ensuremath{\mathparagraph}}
\newcommand{\fmarkvi}{\ensuremath{\|}}
\newcommand{\fmarkvii}{**}
\newcommand{\fmarkviii}{\ensuremath{\dagger\dagger}}
\newcommand{\fmarkix}{\ensuremath{\ddagger\ddagger}}
                
\def\@fnsymbol#1{{\ifcase#1\or \fmarki\or \fmarkii\or \fmarkiii\or \fmarkiv\or \fmarkv\or \fmarkvi\or \fmarkvii\or \fmarkviii\or \fmarkix \else\@ctrerr\fi}}
\makeatother

\renewcommand{\fmarki}{$\dagger$}
\renewcommand{\fmarkii}{*}

\usepackage{graphicx}
\usepackage{dcolumn}
\usepackage{bm}

\begin{document}

\preprint{APS/123-QED}

\title{Zero-Threshold PT-Symmetric Polariton-Raman Laser}

\author{A. Dhara}
\thanks{These authors contributed equally}
\author{P. Das}
\thanks{These authors contributed equally}
\author{D. Chakrabarty}
\thanks{These authors contributed equally}
\affiliation{Department of Physics, Indian Institute of Technology, Kharagpur-721302, India}
\author{K. Ghosh}
\author{A. Roy Chaudhuri}
\affiliation{Materials Science Centre, Indian Institute of Technology, Kharagpur-721302, India}
\author{S. Dhara}
\email{sajaldhara@phy.iitkgp.ac.in}
\affiliation{Department of Physics, Indian Institute of Technology, Kharagpur-721302, India}

\date{\today}

\begin{abstract}
Anisotropy endows topological aspects in optical systems and furnishes a platform to explore non-Hermitian physics, which can be harnessed for the polarization-selective amplification of light. Here, we show a zero-threshold Raman laser can be achieved in an anisotropic optical microcavity via polarization-controlled optical pumping. A loss-gain mechanism between two polarized Stokes modes arises naturally via polarization-dependent stimulated scattering and anisotropic Raman gain of the active layered material inside the microcavity. A Parity-Time (PT) symmetric Hamiltonian has been proposed to explain the emergence of a single polarization mode, essential for achieving a zero-threshold lasing condition. Additionally, intensity correlation measurements of the Stokes modes validate the coherence properties of the emitted light. Our realization of the zero-threshold Raman laser in anisotropic microcavity can open up a new research direction exploring non-Hermitian and topological aspects of light in anisotropic two-dimensional materials.
\end{abstract}

\maketitle

Optical anisotropy in biaxial media introduces a myriad of intriguing phenomena owing to the existence of Hamilton’s diabolical points at the ray-surface which is analogous to the Dirac point in the electronic band structure of graphene. A biaxial medium in addition to complex permittivity possesses singular axis \cite{berryOpticalSingularitiesBirefringent2003} that can offer platforms for exploring non-Hermitian physics, and engineering topological aspects of wave propagation. It was predicted long ago that biaxial media with complex permittivity could selectively amplify circularly polarized light traveling along the singular axis \cite{pancharatnamPropagationLightAbsorbing1955,berryPancharatnamVirtuosoPoincare1994,voigtVIIBehaviourPleochroitic1902}. However, such remarkable effects remain unexplored, possibly due to the challenges of obtaining large single crystals. With the advancement of present technologies and development of two-dimensional (2D) materials, it is now propitious to explore numerous phenomena in anisotropic materials aligned with the growing interest toward understanding non-Hermitian systems in various domains in physics. Recently it has been shown that microcavity polariton systems can be a promising candidate for engineering non-Hermitian topology of polaritonic bands \cite{gaoContinuousTransitionWeak2018, richterExceptionalPointsAnisotropic2017, gaoObservationNonHermitianDegeneracies2015, chakrabartyAnisotropicExcitonPolariton2023, yuen-zhouPlexcitonDiracPoints2016,songRoomtemperaturePolaritonicNonHermitian2021, suDirectMeasurementNonHermitian2021, bergholtzExceptionalTopologyNonHermitian2021}. In this work, we show that a single polarized Raman mode can be amplified inside a microcavity hosting anisotropic exciton-polaritons in 2D materials, enabling us to achieve single mode selectivity which is the key criterion of a thresholdless laser \cite{demartiniAnomalousSpontaneousStimulateddecay1988,yokoyamaRateEquationAnalysis1989,bjorkAnalysisSemiconductorMicrocavity1991,khurginHowPurcellFactor2021}. Ultralow-threshold Raman lasers in semiconductors \cite{spillaneUltralowthresholdRamanLaser2002,grudininUltralowthresholdRamanLasing2007,takahashiMicrometrescaleRamanSilicon2013} are promising for their potential applications as frequency tunable coherent light sources for on-chip photonics. A zero-threshold Raman laser which has remained unattained so far could lead to novel technologies for quantum photonics which are beyond any existing finite threshold Raman laser. One such application would be to realize an efficient quantum frequency converter for non-classical light sources including single photons with frequency tunability range not accessible to present technologies in the field of quantum communications \cite{tyumenevTunableStatepreservingFrequency2022}. In this work we demonstrate a single mode zero-threshold Raman laser and explain our results utilizing the loss-gain mechanism in a non-Hermitian system arising naturally via polarization dependent stimulated Raman process and anisotropic Raman gain of the active layered material inside the microcavity. A non-Hermitian system with equal loss and gain possesses PT symmetry, facilitating several useful and non-trivial functionalities like single-mode or chiral lasing and non-reciprocal light propagation \cite{fengSinglemodeLaserParitytime2014,guoObservation$mathcalPmathcalT$SymmetryBreaking2009,pengChiralModesDirectional2016,kepesidisPTsymmetryBreakingSteady2016,zhangHiddenPTSymmetry2017,fengNonHermitianPhotonicsBased2017,el-ganainyNonHermitianPhysicsPT2018}. 

PT-symmetry in non-Hermitian systems has been realized so far in analogy with coupled two level systems which are spatially separated \cite{liSwitchingMicrocavityPolariton2022,ruterObservationParityTime2010,hodaeiParitytimesymmetricMicroringLasers2014}. Here we propose that a two-level system can also be realized considering the two polarization states of the Stokes modes as the basis ($|\psi_1\rangle$,$|\psi_2\rangle$) of a non-Hermitian Hamiltonian in the following form: 

\begin{equation}
H=\left({|\psi}_1\rangle,|\psi_2\rangle\right)\left(\begin{matrix}\epsilon +i{\delta_1} & g\\ g & \epsilon +i{\delta_2} \end{matrix}\right)\left(\begin{matrix}{\langle\psi}_1|\\{\langle\psi}_2|\end{matrix}\right) 
\label{eq:one}
\end{equation}

In this work we show that the above two-level system can represent a non-Hermitian PT-symmetric system that successfully describes the polarization selective amplification of Raman of an anisotropic two-dimensional material inside optical microcavity. The real part $\epsilon$ of the diagonal elements are the energies, while the positive imaginary parts signify the gain due to stimulated Raman scattering of the coherent pump beam. Here, $g$ is the coupling arising from the off-diagonal Raman gain tensor elements. The eigenvalues of this Hamiltonian are $E^\pm=\ \epsilon + i\hbar\xi_{1,2}$, where $\hbar\xi_{1,2} = (\delta_1+{\delta_2})/2\ \pm\ \sqrt{g^2-\left(\frac{\delta_1+\delta_2}{2}\right)^2}$. In the experiment we measure the intensity of two polarization states $|d_1\rangle$ and $|d_2\rangle$ which are the eigenstates of the above Hamiltonian and demonstrate that a single mode zero-threshold Raman laser is possible in an anisotropic optical microcavity by driving the system into a PT-symmetry broken phase by optical pumping. We utilize the strong Raman emission of layered Rhenium disulphide (ReS\textsubscript{2}) with intriguing optical anisotropy resulting in polarized Stokes modes that can be tunable in resonance with the exciton-polaritons as discussed below. 

\begin{figure}
\includegraphics[scale=0.65]{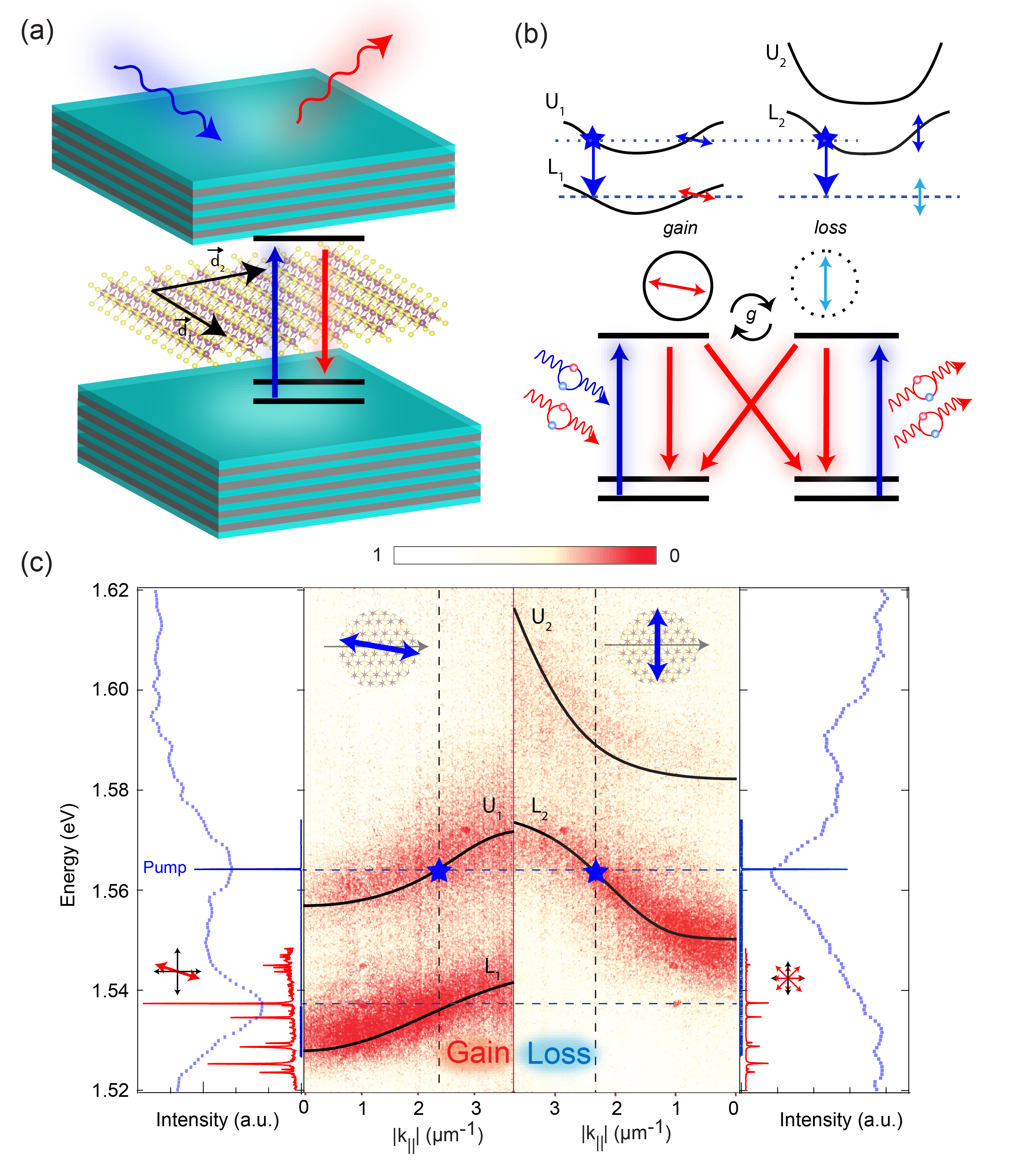}
\caption{\label{fig:1}Schematic and experimental data rendering the mechanism of zero threshold Raman laser. (a) Schematic of Raman scattering in a ReS\textsubscript{2} embedded microcavity, showing the two near-orthogonal directions $d_1$  and $d_2$ where $d_1$ is parallel to the b-axis. (b) Simplified schematic of the stimulated Raman scattering processes involving a pair of polariton dispersions polarized along $d_1$  and $d_2$. The system is pumped at an energy such that both $d_1$ and $d_2$ polarized pump components are resonant with a polariton mode, while the $d_1$ ($d_2$) polarized Stokes-shifted Raman emission coincides with a polariton mode (stop band) thus experiencing gain (loss). (c) Left (right) panel shows angle-resolved reflectance revealing polariton dispersions probed by linearly polarized light along $d_1$ ($d_2$) as indicated in top inset. Extreme left (right) panel shows the Raman spectra for pump polarization $d_1$ ($d_2$), overlaid with the reflectance line plot for the specific $k_{||}$ where pump energy is in resonance with the U\textsubscript{1} and L\textsubscript{2} polariton modes, indicated with a dashed line in the angle resolved data. Pump laser spectrum is shown in blue (intensity in arbitrary scale).}
\end{figure}

Two dimensional semiconductors like ReS\textsubscript{2} are being recognized as a potential active laser medium and an ideal platform for realising exciton-polaritons \cite{wuMonolayerSemiconductorNanocavity2015, yeMonolayerExcitonicLaser2015, shangRoomtemperature2DSemiconductor2017, liuStrongLightMatter2015, dharaAnomalousDispersionMicrocavity2018}. ReS\textsubscript{2} is known for its distorted 1T crystal structure hosting strongly polarized excitons  \cite{aslanLinearlyPolarizedExcitons2016, dharaAdditionalExcitonicFeatures2020}, with tunable light-matter coupling \cite{chakrabartyInterfacialAnisotropicExcitonpolariton2021}. The ReS\textsubscript{2}-embedded microcavity supports two sets of non-orthogonal exciton-polariton modes with polarization tunable dispersions. Such systems are being understood recently using non-Hermitian descriptions \cite{pengParityTimeSymmetric2014,changParityTimeSymmetry2014,gaoChiralModesExceptional2018,zhangPhononLaserOperating2018,kremerDemonstrationTwodimensionalPTsymmetric2019,miriExceptionalPointsOptics2019,ozdemirParityTimeSymmetry2019,khurginExceptionalPointsPolaritonic2020}. Due to reduced crystal symmetry, ReS\textsubscript{2} has 18 Raman modes in the range of 100 to 450 cm\textsuperscript{-1} \cite{mccrearyIntricateResonantRaman2017, pradhanMetalInsulatorQuantumPhase2015, fengRamanVibrationalSpectra2015}. Interestingly we observed the presence of strong Raman peaks and vanishing photoluminescence at near resonance excitation similar to what has been reported recently in ReS\textsubscript{2} \cite{jadczakExcitonBindingEnergy2019}, which makes it a unique platform to observe Raman lasing. The ReS\textsubscript{2}-microcavity polaritonic system can thus be pumped to achieve polarization dependent stimulated resonant Raman scattering \cite{fainsteinRamanScatteringEnhancement1995, fainsteinCavityPolaritonMediatedResonant1997}. A $\sim$10 nm thick ReS\textsubscript{2} crystal is mechanically exfoliated and placed in an optical microcavity (see Methods in Supplemental Material \cite{SeeSupplementalMaterial} for fabrication details), as illustrated in Fig.\ \ref{fig:1}(a). The absorption of two excitonic oscillators (X\textsubscript{1} and X\textsubscript{2}) in ReS\textsubscript{2} show high polarization dependence due to reduced crystal symmetry of the lattice resulting in the biaxiality with complex permittivity. Strong light-matter interactions inside optical microcavity, leads to the formation of exciton-polariton pairs, as shown in Fig.\ \ref{fig:1}(b,c) (see Note S1 in Supplemental Material \cite{SeeSupplementalMaterial} for cavity parameters). The intricate topology of these exciton-polariton bands are investigated in great detail which is reported in another work \cite{chakrabartyAnisotropicExcitonPolariton2023}. We define two non-orthogonal polarization directions with respect to the crystallographic b-axis, 170\textsuperscript{o} ($d_1$)  and 90\textsuperscript{o} $(d_2$) corresponding to the absorption maxima attributed to X\textsubscript{1} and X\textsubscript{2} polariton branches respectively. The four resulting polariton modes are designated as two upper (U\textsubscript{1}, U\textsubscript{2}), and two lower (L\textsubscript{1}, L\textsubscript{2}) polariton branches. We choose the pump beam energy at 1.564 eV which is in resonance with both the U\textsubscript{1} and L\textsubscript{2} branches, to get equal amplification of the pump beam for both the polarization modes inside the microcavity. The schematic in the lower panel in Fig.\ \ref{fig:1}(b) depicts the microscopic mechanism of the stimulated Raman process inside microcavity giving rise to the relative gain and loss for $d_1$  and $d_2$  polarized Stokes modes. 

\begin{figure*}
\includegraphics[scale=1]{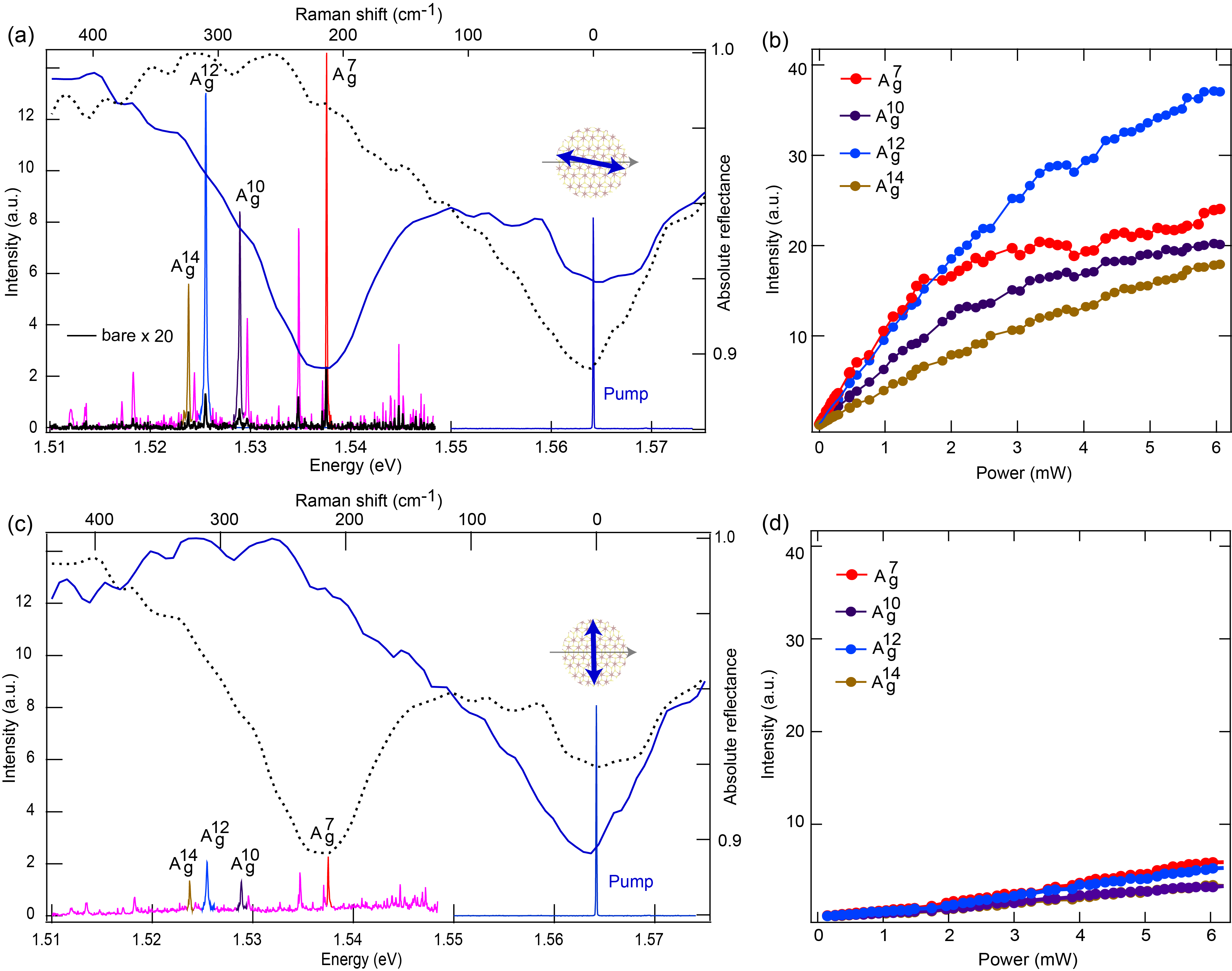}
\caption{\label{fig:2}Raman laser power dependence at 4 K. (a) Raman spectrum for $d_1$ polarized pump, overlaid with $d_1$ (solid line) and $d_2$ (dotted line) polarized reflectance line plot from Fig 1(c). Four distinct Raman modes are labelled with different colours. For comparison, the Raman spectrum from bare ReS\textsubscript{2}/SiO\textsubscript{2} (black, multiplied 20X) is also shown. (b) Power dependence of $d_1$ component of Raman intensity for $d_1$ polarized pump, showing linear dependence followed by saturation at higher power. (c, d) Same as (a, b) for $d_2$ component of Raman intensity with $d_2$  polarized pump.}
\end{figure*}

Polarization and angle resolved reflectivity measured at 4 K (see Methods in Supplemental Material  \cite{SeeSupplementalMaterial} and reference \cite{pattanayakSteadyStateApproachStudying2022} therein) reveals the four polariton modes, as shown in Fig.\ \ref{fig:1}(c) where the crystallographic b-axis of ReS\textsubscript{2} is oriented along $k_{||}$. Plot in the extreme left panel in Fig.\ \ref{fig:1}(c) shows amplified stimulated Raman emission when the pump beam is polarized along $d_1$ and the Stokes lines lie within the L\textsubscript{1} polariton band, thus satisfying a double resonance condition. The Raman intensity reduces drastically when the pump is polarized along $d_2$ and the Stokes lines lies in the stop band of the cavity, as shown in the right extreme panel in Fig. \ref{fig:1}(c). To understand the nature of the amplified Raman signals we first measure the power dependence of the Raman spectrum, as discussed below. Observed Raman modes are assigned based on existing literature  \cite{mccrearyIntricateResonantRaman2017, pradhanMetalInsulatorQuantumPhase2015, zhouStackingOrderDrivenOpticalProperties2020} of ReS\textsubscript{2} with AA stacking order (see Note S2, Fig. S2-S3 and Table S2 in Supplemental Material \cite{SeeSupplementalMaterial} for characterization of the Raman modes). We pick four representative Raman modes, labeled with different colours: $A_g^7$  (214 cm\textsuperscript{-1}), $A_g^{10}$ (284 cm\textsuperscript{-1}), $A_g^{12}$ (311 cm\textsuperscript{-1}) and $A_g^{14}$ (325 cm\textsuperscript{-1}) as shown in Fig. \ref{fig:2}(a)-\ref{fig:2}(c). We find the Raman emission is amplified by $~\sim$150-300 times compared to bare ReS\textsubscript{2} when the pump is polarized along $d_1$ (shown in Fig. \ref{fig:2}a). The spectral linewidths ($\sim$0.8 cm\textsuperscript{-1}) of these Raman peaks are within the spectrometer resolution. The amplified Raman intensity shows a linear power dependence (down to 180 nW, see Supplementary Material Fig. S4) which gets saturated at higher power, revealing the signature of zero-threshold lasing as shown in Fig. \ref{fig:2}(b). A similar power dependence was reported earlier in trapped single atom laser from microcavity \cite{mckeeverExperimentalRealizationOneatom2003}. The coherence characteristics of the Raman laser has been verified with the second order correlation ($g^{\left(2\right)}\left(\tau\right)$) measurement which is discussed in later section. 

\begin{figure*}
\includegraphics[scale=0.95]{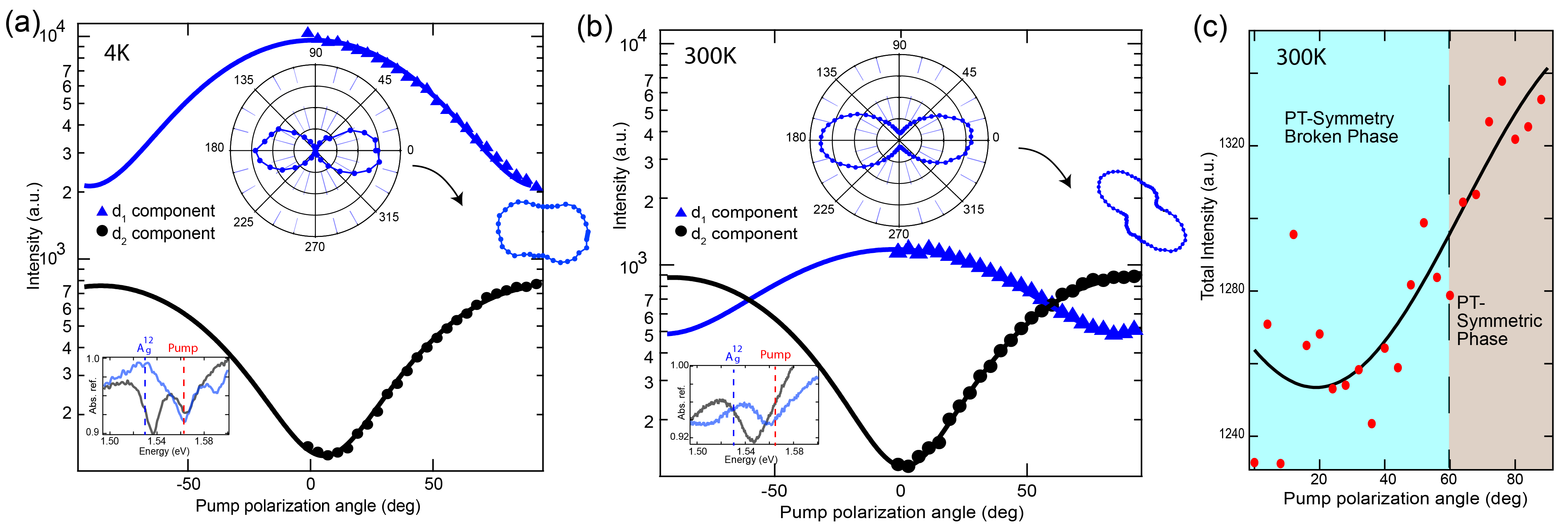}
\caption{\label{fig:3}Pump polarization dependence of the $d_1$ and $d_2$ components of $A_g^{12}$ mode intensity at (a) 4K and (b) room temperature (RT), shown in logarithmic scale. (Middle and right insets) Polar plots showing $A_g^{12}$ mode intensity with analyzer angle w.r.t. b-axis, for $d_1$ and $d_2$ polarized pump. Solid curves are obtained from the microscopic theory. Bottom-left inset plots indicate the relative position of the pump and emission energies with the cavity modes at 4K and RT. (c) Pump polarization dependence of total intensity variation revealing a PT symmetric phase transition.}
\end{figure*}

As we have described earlier the observation of zero-threshold lasing can be understood from the loss-gain mechanism in a non-Hermitian system. When the pump beam is $d_1$  polarized the Stokes modes are in resonance with the lower polariton L\textsubscript{1}, thus experiencing amplification due to Purcell enhancement. Conversely, for $d_2$  polarized pump, the Stokes modes fall within the stop band of the cavity, thus experiencing relative loss resulting in reduced Raman intensity as shown in Fig. \ref{fig:2}(c)-(d). Hence, the observed zero-threshold lasing can be understood considering the two-state system governed by a non-Hermitian PT-symmetric Hamiltonian as described above in Eq. \ref{eq:one}. A detailed microscopic theory considering stimulated Raman scattering inside microcavity \cite{yarivQuantumElectronics1989,wuTheoryMicrocavityenhancedRaman1999} has been developed to directly obtain the transition rates which can be identified as the imaginary part of the eigenvalues ($\xi_1$, $\xi_2$) of Eq. \ref{eq:one} (see Note S3 in the Supplemental Material \cite{SeeSupplementalMaterial}):

\begin{multline*}
    \xi_j=F_j{[g}_{jj}Q_jI_0\cos\left(\phi_j-\theta\right)^2+g_{kj}Q_kI_0\cos\left(\phi_k-\theta\right)^2 \\
    -{(g}_{jj}+g_{jk})f(T)]
\end{multline*}

where, $j$ and $k$ correspond to the indices 1, 2. $F_j$ is the Purcell factor for the $j^{th}$ mode at Stokes frequency, $Q_j$ is the amplification factor for the pump, $g_{jj}$ and $g_{kj}$\ are the diagonal and off-diagonal elements of Raman gain tensor of the material, and $f\left(T\right)=\frac{1-\exp{\left(-\frac{\hbar\omega}{2kT}\right)}}{\exp{\left(\frac{\hbar\omega}{kT}\right)}-1}$ where, $\omega$ is the angular frequency of the phonon mode. From our experimental setup and sample orientation we find $\phi_1=0^o$, and $\phi_2={80}^o$. For a particular choice of basis states, $\left|\psi_j\right\rangle=\sum_{k=1}^{2}{{C_k|d}_k\rangle\ }$ one can obtain the parameters $g$, $\delta_1$ and $\delta_2$ from the above expression of $\xi_1$ and $\xi_2$. We estimated the values of Purcell factors $F_1=53$ and $F_2=1.6$ from the quality factors and detuning of the cavity for mode 1 and 2 respectively. The expression for $\xi_j$ obtained from the microscopic theory is further verified by measuring the polarization dependent Raman intensity and the state of polarization of emission. In Fig. \ref{fig:3}(a) (at 4 K) and Fig. \ref{fig:3}(b) (at 300 K) we plot the intensity components of the Raman signal measured along the $d_1$ and $d_2$ directions with the variation of polarization of the pump beam ($\theta$). At $\theta=0^o$ ($d_1$ pump polarization), the emission intensity is highest for the $|d_1\rangle$ mode exhibiting a strong degree of polarization $>$ 0.95 with major axis of intensity profile (measured by analyzer rotation) aligned along $d_1$ polarization (see Fig. 3a, Middle inset). A large difference between $\xi_1$ and $\xi_2$ with $\xi_1\gg\xi_2$,  ensures amplification of the $|d_1\rangle$ mode and a simultaneous abatement of the $|d_2\rangle$ mode, resulting in a highly polarized single mode laser which occurs at the PT-symmetry broken phase for $d_1$ pump polarization. In this regime, the amplification of a single polarized mode ensures that the the value of spectral overlap factor  $\beta \to 1$, meeting the condition for zero-threshold lasing  \cite{yamamotoOpticalProcessesMicrocavities1993, khajavikhanThresholdlessNanoscaleCoaxial2012, prietoThresholdlessLaserOperation2015,wuZerothresholdOpticalGain2017,jagschQuantumOpticalStudy2018,checouryDeterministicMeasurementPurcell2010, petrakPurcellenhancedRamanScattering2014, kavokinMicrocavities2017} which explains the linear power dependence of Raman laser before the saturation at higher pump power  (see Note S3(b) in the Supplemental Material \cite{SeeSupplementalMaterial}).

As the pump polarization rotates from $d_1$ to $d_2$  ($\theta\sim90^o$), the overall Raman emission intensity of $d_1$ mode reduces but it remains much higher than the $d_2$ mode. What is more counterintuitive is that the degree of polarization reduces significantly to $<$0.4 at $\theta\sim90^o$ as shown in the right inset in Fig. \ref{fig:3}(a). The same $\theta$ variation of $d_1$ and $d_2$ components of the Raman mode show very different characteristics at 300 K as shown in Fig. \ref{fig:3}(b), where the intensities of $d_1$ and $d_2$ components cross each other at $\theta\sim60^o$. We show below that such counterintuitive polarization variation of the $d_1$ and $d_2$ components can be understood in the framework of non-Hermitian PT-symmetry. According to our model, when the pump beam is $d_2$ polarized, the system is still in the PT-symmetry broken phase but the inequality $\xi_1>\xi_2$ become relatively weaker resulting in unpolarized emission of the two Stokes modes.  Importantly, the validity of our model is verified by predicting the data at 300 K quite accurately as shown in Fig. \ref{fig:3}(b) where similar ratio of Raman gain tensor elements $g_{jj}$, $g_{kj}$ obtained at 4 K have been utilized with suitable modification of the Purcell factors as the cavity is detuned. Interestingly, we observe that at T=300 K, the system can reach to the PT-symmetric condition at particular $\theta=90^o$ When $\xi_1 = \xi_2$ which corresponds to the crossing of intensities of the $d_1$ and $d_2$ components as seen in Fig. \ref{fig:3}(b). To better understand the phase transition at room temperature, we plot the total intensity variation with respect to pump polarization. Initially, the total intensity decreases as the gain contrast reduces; however, intriguingly, the trend reverses after $\theta=20^o$. The experimental data is comparable with the theory as illustrated in Fig. \ref{fig:3}(c). This non-trivial phenomenon is analogous to loss-induced transparency reported in passive PT-symmetric systems \cite{guoObservation$mathcalPmathcalT$SymmetryBreaking2009}. 

\begin{figure}
\includegraphics[scale=0.9]{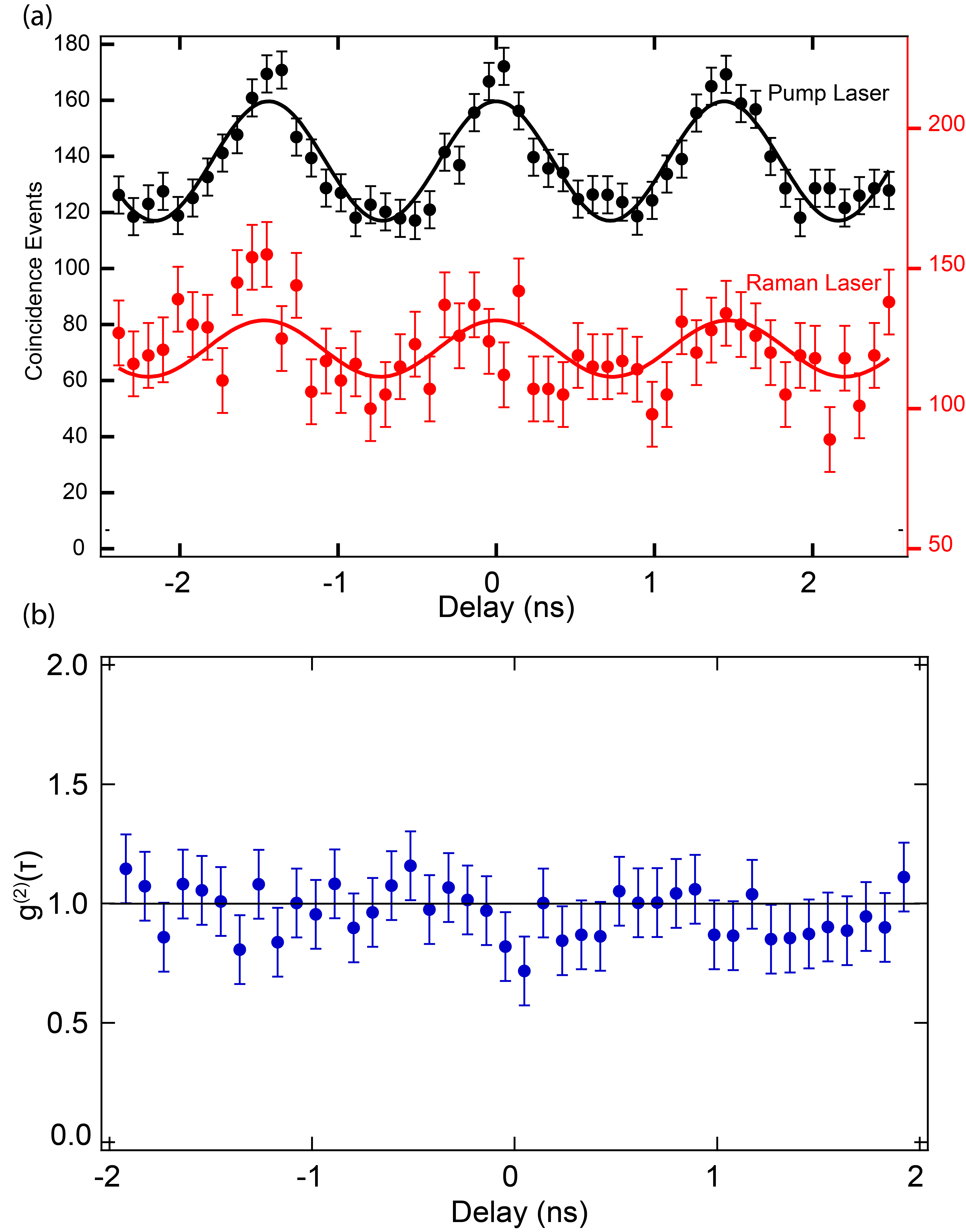}
\caption{\label{fig:4}Second-order correlation $g^{\left(2\right)}(\tau)$ measurements of the zero-threshold Raman laser.  (a) Coincidence events recorded using a Hanbury-Brown-Twiss (HBT) setup for the reflected pump laser (black circles) and $A_g^{12}$ Raman mode emission (red circles) at higher power with $d_1$ polarized pump. The solid lines show best fits obtained from the model. (b) $g^{\left(2\right)}(\tau)$  for the $A_g^{12}$ Raman mode obtained by normalizing coincidence events with respect to the pump laser. The error bars indicate the uncertainty due to Poisson distribution statistics.}
\end{figure}

Second-order correlation ($g^{\left(2\right)}$) measurements were performed to characterize the coherence property of the Raman laser. The Raman mode with the strongest emission ($A_g^{12}$) was selected for the $g^{\left(2\right)}$ measurement. We first characterized the CW pump laser reflected from our sample and found a slight sinusoidal modulation in $g^{\left(2\right)}$ (see Fig. \ref{fig:4}(a)) which is signifying the presence of satellite peaks at frequencies $\omega_0\pm\Delta\omega$ in the spectrum of the reflected pump beam centred at $\omega_0$. The value of $\Delta\omega$ was obtained as $\sim$2 GHz (see Methods in Supplemental Material \cite{SeeSupplementalMaterial}). After characterizing the $g^{\left(2\right)}$ of the reflected pump beam, we measured the $g^{\left(2\right)}$  for $A_g^{12}$ mode at $>$6 mW pump power. Figure \ref{fig:4}(a) shows the result of $g^{\left(2\right)}$ measurements in red, where we do not observe the peak at $g^{\left(2\right)}(\tau=0)$  expected for spontaneous emission. Instead, $g^{\left(2\right)}(\tau)$  retains the variation observed for the reflected pump beam, which provides further proof of coherence of the emitted Raman laser signal. After normalizing with the $g^{\left(2\right)}$ of the pump laser, we obtain a flat $g^{\left(2\right)}$ which is the signature of a laser as shown in Fig. \ref{fig:4}(b). 

\begin{figure*}
\includegraphics[scale=0.95]{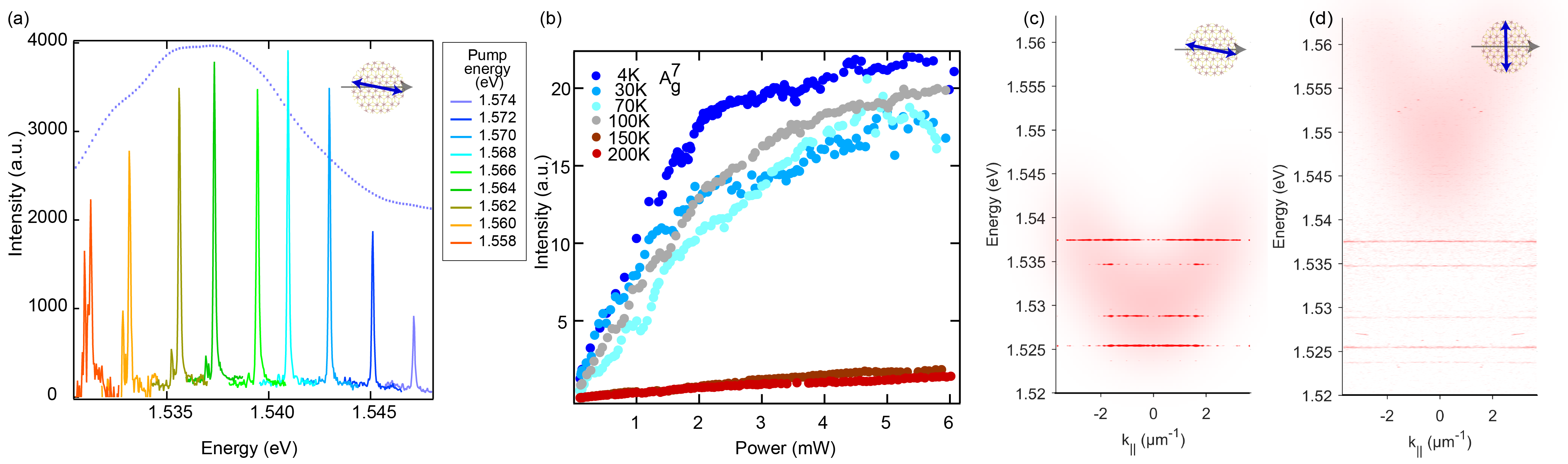}
\caption{\label{fig:5}Frequency tunability and temperature range along with directionality of the Raman Laser: (a) Intensity variation of the $A_g^7$ mode spectrum for $d_1$-polarized pump, with different colors representing different pump energies. The reflectance for $d_1$-polarization from Fig 2(a) has been inverted and overlaid as a visual aid. (b) Evolution of the power dependence with temperature for $A_g^7$  mode with $d_1$ polarized pump. (c) Angle-resolved Raman spectrum averaged over positive and negative $k_{||}$, displaying the directionality of emission for $d_1$ polarized pump, with polariton band L\textsubscript{1} indicated by the translucent red region in the background. (d) Same as (c) for $d_2$ polarized pump, with the L\textsubscript{2} polariton band highlighted. The same Raman modes do not lase or display any emission pattern.}
\end{figure*}

The frequency tunability of the Raman laser is investigated and shown in Fig. \ref{fig:5}(a). The intensity of the $A_g^7$  lasing mode as a function of pump energy for $d_1$-polarized excitation closely follows the reflectance dip of the L\textsubscript{1} polariton branch (the reflectance is inverted and overlaid in Fig. \ref{fig:5}(a) as a guide to the eye). For $d_2$-polarized excitation the same mode does not show such pronounced intensity variation (see Fig. S5 in Supplemental Material \cite{SeeSupplementalMaterial}). The operational temperature range for the zero-threshold laser, and the transition to non-lasing mode has been investigated in the temperature dependent power variation, as shown in Fig. \ref{fig:5}(b). We observe that the characteristic linear power dependence with high amplification followed by saturation at high pump power, a distinctive feature of zero-threshold lasing, remains robust up to 100K. However, beyond this temperature, the emission intensity drops sharply, indicating the double resonance condition is no longer fulfilled ($A_g^{12}$ mode shows the same temperature-dependence, see Fig.\ S6 in Supplemental Material \cite{SeeSupplementalMaterial}). Consequently, the lasing condition $\xi_1\gg\xi_2$ as discussed above is not satisfied at temperatures above 100K. The double resonance condition is also responsible for directional emission, a fundamental laser property, as shown by enhanced intensity distribution of Raman modes ($A_g^7$, $A_g^{10}$, $A_g^{12}$ and $A_g^{14}$) coinciding with the polariton band (shown in the shaded background) in Fig.\ \ref{fig:5}(c) under  $d_1$ excitation. In contrast, $d_2$ excitation produces an isotropic emission pattern, indicating non-lasing behavior, as shown in Fig.\ \ref{fig:5}(d). The robustness of zero threshold lasing and reproducibility of all the experimental results as discussed above has been observed in a different ReS\textsubscript{2} flake of similar thickness inside the same cavity, kept at the same orientation as the previous one, which is shown in Fig. S7 (see Supplemental Material \cite{SeeSupplementalMaterial}). 

In conclusion, we utilize anisotropic 2D materials inside a microcavity to demonstrate a non-trivial phenomenon of zero-threshold Raman lasing. We propose a non-Hermitian PT symmetric mechanism involving loss and gain of two Stokes polarization modes which successfully explains our observation of highly polarized zero-threshold lasing. The realization of a zero-threshold Raman laser may motivate further research in exploring other optical phenomena involving resonant nonlinear and parametric processes in anisotropic 2D materials in optical microcavities. Thus, our work can pave the way to a new research direction involving non-Hermitian PT-symmetry and topological aspects of wave propagation in anisotropic media which can lead to a significant progress in quantum technology as well as it can provide a platform for fundamental research in condensed matter and quantum optics.

\begin{acknowledgments}
SD conceptualized the project. AD, DC, KG, ARC and SD contributed in the sample fabrication. AD, DC, PD and SD formulated the experiments, performed data analysis. PD, DC and SD developed the theory. AD, DC, PD performed all optical measurements. AD, DC, PD and SD drafted the paper, and all authors contributed to reviewing and editing the final draft. SD supervised the project. This work has been supported by funding from the Science and Engineering Research Board (CRG/2018/002845, CRG/2021/000811); Ministry of Education (MoE/STARS- 1/647); Council of Scientific and Industrial Research, India (09/081(1352)/2019-EMR-I); Department of Science and Technology (IN), INSPIRE
Fellowship (IF180046) and Indian Institute of Technology Kharagpur. We thank M. V. Berry, S. P. Khastgir, G. S. Agarwal, J. Bhattacharya and T. Chakraborty for valuable discussions on this work. 
\end{acknowledgments}

\begin{thebibliography}{66}%
\makeatletter
\providecommand \@ifxundefined [1]{%
 \@ifx{#1\undefined}
}%
\providecommand \@ifnum [1]{%
 \ifnum #1\expandafter \@firstoftwo
 \else \expandafter \@secondoftwo
 \fi
}%
\providecommand \@ifx [1]{%
 \ifx #1\expandafter \@firstoftwo
 \else \expandafter \@secondoftwo
 \fi
}%
\providecommand \natexlab [1]{#1}%
\providecommand \enquote  [1]{``#1''}%
\providecommand \bibnamefont  [1]{#1}%
\providecommand \bibfnamefont [1]{#1}%
\providecommand \citenamefont [1]{#1}%
\providecommand \href@noop [0]{\@secondoftwo}%
\providecommand \href [0]{\begingroup \@sanitize@url \@href}%
\providecommand \@href[1]{\@@startlink{#1}\@@href}%
\providecommand \@@href[1]{\endgroup#1\@@endlink}%
\providecommand \@sanitize@url [0]{\catcode `\\12\catcode `\$12\catcode `\&12\catcode `\#12\catcode `\^12\catcode `\_12\catcode `\%12\relax}%
\providecommand \@@startlink[1]{}%
\providecommand \@@endlink[0]{}%
\providecommand \url  [0]{\begingroup\@sanitize@url \@url }%
\providecommand \@url [1]{\endgroup\@href {#1}{\urlprefix }}%
\providecommand \urlprefix  [0]{URL }%
\providecommand \Eprint [0]{\href }%
\providecommand \doibase [0]{https://doi.org/}%
\providecommand \selectlanguage [0]{\@gobble}%
\providecommand \bibinfo  [0]{\@secondoftwo}%
\providecommand \bibfield  [0]{\@secondoftwo}%
\providecommand \translation [1]{[#1]}%
\providecommand \BibitemOpen [0]{}%
\providecommand \bibitemStop [0]{}%
\providecommand \bibitemNoStop [0]{.\EOS\space}%
\providecommand \EOS [0]{\spacefactor3000\relax}%
\providecommand \BibitemShut  [1]{\csname bibitem#1\endcsname}%
\let\auto@bib@innerbib\@empty
\bibitem [{\citenamefont {Berry}\ and\ \citenamefont {Dennis}(2003)}]{berryOpticalSingularitiesBirefringent2003}%
  \BibitemOpen
  \bibfield  {author} {\bibinfo {author} {\bibfnamefont {M.~V.}\ \bibnamefont {Berry}}\ and\ \bibinfo {author} {\bibfnamefont {M.~R.}\ \bibnamefont {Dennis}},\ }\bibfield  {title} {\bibinfo {title} {The optical singularities of birefringent dichroic chiral crystals},\ }\href {https://doi.org/10.1098/rspa.2003.1155} {\bibfield  {journal} {\bibinfo  {journal} {Proceedings of the Royal Society of London. Series A: Mathematical, Physical and Engineering Sciences}\ }\textbf {\bibinfo {volume} {459}},\ \bibinfo {pages} {1261} (\bibinfo {year} {2003})}\BibitemShut {NoStop}%
\bibitem [{\citenamefont {Pancharatnam}(1955)}]{pancharatnamPropagationLightAbsorbing1955}%
  \BibitemOpen
  \bibfield  {author} {\bibinfo {author} {\bibfnamefont {S.}~\bibnamefont {Pancharatnam}},\ }\bibfield  {title} {\bibinfo {title} {The propagation of light in absorbing biaxial crystals --- {{I}}. {{Theoretical}}},\ }\href {https://doi.org/10.1007/BF03053496} {\bibfield  {journal} {\bibinfo  {journal} {Proceedings of the Indian Academy of Sciences - Section A}\ }\textbf {\bibinfo {volume} {42}},\ \bibinfo {pages} {86} (\bibinfo {year} {1955})}\BibitemShut {NoStop}%
\bibitem [{\citenamefont {Berry}(1994)}]{berryPancharatnamVirtuosoPoincare1994}%
  \BibitemOpen
  \bibfield  {author} {\bibinfo {author} {\bibfnamefont {M.}~\bibnamefont {Berry}},\ }\bibfield  {title} {\bibinfo {title} {Pancharatnam, virtuoso of the {{Poincar{\'e}}} sphere: An appreciation},\ }\href{https://www.jstor.org/stable/24095727} {\bibfield  {journal} {\bibinfo  {journal} {Current Science}\ }\textbf  {\bibinfo {volume} {67}},\ \bibinfo {pages} {220} (\bibinfo {year} {1994})} \textbf{}\  \BibitemShut{NoStop}%
\bibitem [{\citenamefont {Voigt}(1902)}]{voigtVIIBehaviourPleochroitic1902}%
  \BibitemOpen
  \bibfield  {author} {\bibinfo {author} {\bibfnamefont {W.}~\bibnamefont {Voigt}},\ }\bibfield  {title} {\bibinfo {title} {{{VII}}. {{On}} the behaviour of pleochroitic crystals along directions in the neighbourhood of an optic axis},\ }\href {https://doi.org/10.1080/14786440209462820} {\bibfield  {journal} {\bibinfo  {journal} {The London, Edinburgh, and Dublin Philosophical Magazine and Journal of Science}\ }\textbf {\bibinfo {volume} {4}},\ \bibinfo {pages} {90} (\bibinfo {year} {1902})}\BibitemShut {NoStop}%
\bibitem [{\citenamefont {Gao}\ \emph {et~al.}(2018{\natexlab{a}})\citenamefont {Gao}, \citenamefont {Li}, \citenamefont {Bamba},\ and\ \citenamefont {Kono}}]{gaoContinuousTransitionWeak2018}%
  \BibitemOpen
  \bibfield  {author} {\bibinfo {author} {\bibfnamefont {W.}~\bibnamefont {Gao}}, \bibinfo {author} {\bibfnamefont {X.}~\bibnamefont {Li}}, \bibinfo {author} {\bibfnamefont {M.}~\bibnamefont {Bamba}},\ and\ \bibinfo {author} {\bibfnamefont {J.}~\bibnamefont {Kono}},\ }\bibfield  {title} {\bibinfo {title} {Continuous transition between weak and ultrastrong coupling through exceptional points in carbon nanotube microcavity exciton--polaritons},\ }\href {https://doi.org/10.1038/s41566-018-0157-9} {\bibfield  {journal} {\bibinfo  {journal} {Nature Photonics}\ }\textbf {\bibinfo {volume} {12}},\ \bibinfo {pages} {362} (\bibinfo {year} {2018}{\natexlab{a}})}\BibitemShut {NoStop}%
\bibitem [{\citenamefont {Richter}\ \emph {et~al.}(2017)\citenamefont {Richter}, \citenamefont {Michalsky}, \citenamefont {Sturm}, \citenamefont {Rosenow}, \citenamefont {Grundmann},\ and\ \citenamefont {{Schmidt-Grund}}}]{richterExceptionalPointsAnisotropic2017}%
  \BibitemOpen
  \bibfield  {author} {\bibinfo {author} {\bibfnamefont {S.}~\bibnamefont {Richter}}, \bibinfo {author} {\bibfnamefont {T.}~\bibnamefont {Michalsky}}, \bibinfo {author} {\bibfnamefont {C.}~\bibnamefont {Sturm}}, \bibinfo {author} {\bibfnamefont {B.}~\bibnamefont {Rosenow}}, \bibinfo {author} {\bibfnamefont {M.}~\bibnamefont {Grundmann}},\ and\ \bibinfo {author} {\bibfnamefont {R.}~\bibnamefont {{Schmidt-Grund}}},\ }\bibfield  {title} {\bibinfo {title} {Exceptional points in anisotropic planar microcavities},\ }\href {https://doi.org/10.1103/PhysRevA.95.023836} {\bibfield  {journal} {\bibinfo  {journal} {Physical Review A}\ }\textbf {\bibinfo {volume} {95}},\ \bibinfo {pages} {023836} (\bibinfo {year} {2017})}\BibitemShut {NoStop}%
\bibitem [{\citenamefont {Gao}\ \emph {et~al.}(2015)\citenamefont {Gao}, \citenamefont {Estrecho}, \citenamefont {Bliokh}, \citenamefont {Liew}, \citenamefont {Fraser}, \citenamefont {Brodbeck}, \citenamefont {Kamp}, \citenamefont {Schneider}, \citenamefont {H{\"o}fling}, \citenamefont {Yamamoto}, \citenamefont {Nori}, \citenamefont {Kivshar}, \citenamefont {Truscott}, \citenamefont {Dall},\ and\ \citenamefont {Ostrovskaya}}]{gaoObservationNonHermitianDegeneracies2015}%
  \BibitemOpen
  \bibfield  {author} {\bibinfo {author} {\bibfnamefont {T.}~\bibnamefont {Gao}}, \bibinfo {author} {\bibfnamefont {E.}~\bibnamefont {Estrecho}}, \bibinfo {author} {\bibfnamefont {K.~Y.}\ \bibnamefont {Bliokh}}, \bibinfo {author} {\bibfnamefont {T.~C.~H.}\ \bibnamefont {Liew}}, \bibinfo {author} {\bibfnamefont {M.~D.}\ \bibnamefont {Fraser}}, \bibinfo {author} {\bibfnamefont {S.}~\bibnamefont {Brodbeck}}, \bibinfo {author} {\bibfnamefont {M.}~\bibnamefont {Kamp}}, \bibinfo {author} {\bibfnamefont {C.}~\bibnamefont {Schneider}}, \bibinfo {author} {\bibfnamefont {S.}~\bibnamefont {H{\"o}fling}}, \bibinfo {author} {\bibfnamefont {Y.}~\bibnamefont {Yamamoto}}, \bibinfo {author} {\bibfnamefont {F.}~\bibnamefont {Nori}}, \bibinfo {author} {\bibfnamefont {Y.~S.}\ \bibnamefont {Kivshar}}, \bibinfo {author} {\bibfnamefont {A.~G.}\ \bibnamefont {Truscott}}, \bibinfo {author} {\bibfnamefont {R.~G.}\ \bibnamefont {Dall}},\ and\ \bibinfo {author} {\bibfnamefont {E.~A.}\ \bibnamefont {Ostrovskaya}},\ }\bibfield  {title}
  {\bibinfo {title} {Observation of non-{{Hermitian}} degeneracies in a chaotic exciton-polariton billiard},\ }\href {https://doi.org/10.1038/nature15522} {\bibfield  {journal} {\bibinfo  {journal} {Nature}\ }\textbf {\bibinfo {volume} {526}},\ \bibinfo {pages} {554} (\bibinfo {year} {2015})}\BibitemShut {NoStop}%
\bibitem [{\citenamefont {Chakrabarty}\ \emph {et~al.}(2023)\citenamefont {Chakrabarty}, \citenamefont {Dhara}, \citenamefont {Das}, \citenamefont {Ghosh}, \citenamefont {Chaudhuri},\ and\ \citenamefont {Dhara}}]{chakrabartyAnisotropicExcitonPolariton2023}%
  \BibitemOpen
  \bibfield  {author} {\bibinfo {author} {\bibfnamefont {D.}~\bibnamefont {Chakrabarty}}, \bibinfo {author} {\bibfnamefont {A.}~\bibnamefont {Dhara}}, \bibinfo {author} {\bibfnamefont {P.}~\bibnamefont {Das}}, \bibinfo {author} {\bibfnamefont {K.}~\bibnamefont {Ghosh}}, \bibinfo {author} {\bibfnamefont {A.~R.}\ \bibnamefont {Chaudhuri}},\ and\ \bibinfo {author} {\bibfnamefont {S.}~\bibnamefont {Dhara}},\ }\href {https://doi.org/10.48550/arXiv.2305.17472} {\bibinfo {title} {Anisotropic exciton polariton pairs as a platform for {{PT-symmetric}} non-{{Hermitian}} physics}} (\bibinfo {year} {2023}),\ \Eprint {https://arxiv.org/abs/2305.17472} {arXiv:2305.17472 [cond-mat, physics:physics, physics:quant-ph]} \BibitemShut {NoStop}%
\bibitem [{\citenamefont {{Yuen-Zhou}}\ \emph {et~al.}(2016)\citenamefont {{Yuen-Zhou}}, \citenamefont {Saikin}, \citenamefont {Zhu}, \citenamefont {Onbasli}, \citenamefont {Ross}, \citenamefont {Bulovic},\ and\ \citenamefont {Baldo}}]{yuen-zhouPlexcitonDiracPoints2016}%
  \BibitemOpen
  \bibfield  {author} {\bibinfo {author} {\bibfnamefont {J.}~\bibnamefont {{Yuen-Zhou}}}, \bibinfo {author} {\bibfnamefont {S.~K.}\ \bibnamefont {Saikin}}, \bibinfo {author} {\bibfnamefont {T.}~\bibnamefont {Zhu}}, \bibinfo {author} {\bibfnamefont {M.~C.}\ \bibnamefont {Onbasli}}, \bibinfo {author} {\bibfnamefont {C.~A.}\ \bibnamefont {Ross}}, \bibinfo {author} {\bibfnamefont {V.}~\bibnamefont {Bulovic}},\ and\ \bibinfo {author} {\bibfnamefont {M.~A.}\ \bibnamefont {Baldo}},\ }\bibfield  {title} {\bibinfo {title} {Plexciton {{Dirac}} points and topological modes},\ }\href {https://doi.org/10.1038/ncomms11783} {\bibfield  {journal} {\bibinfo  {journal} {Nature Communications}\ }\textbf {\bibinfo {volume} {7}},\ \bibinfo {pages} {11783} (\bibinfo {year} {2016})}\BibitemShut {NoStop}%
\bibitem [{\citenamefont {Song}\ \emph {et~al.}(2021)\citenamefont {Song}, \citenamefont {Choi}, \citenamefont {Woo}, \citenamefont {Park},\ and\ \citenamefont {Cho}}]{songRoomtemperaturePolaritonicNonHermitian2021}%
  \BibitemOpen
  \bibfield  {author} {\bibinfo {author} {\bibfnamefont {H.~G.}\ \bibnamefont {Song}}, \bibinfo {author} {\bibfnamefont {M.}~\bibnamefont {Choi}}, \bibinfo {author} {\bibfnamefont {K.~Y.}\ \bibnamefont {Woo}}, \bibinfo {author} {\bibfnamefont {C.~H.}\ \bibnamefont {Park}},\ and\ \bibinfo {author} {\bibfnamefont {Y.-H.}\ \bibnamefont {Cho}},\ }\bibfield  {title} {\bibinfo {title} {Room-temperature polaritonic non-{{Hermitian}} system with single microcavity},\ }\href {https://doi.org/10.1038/s41566-021-00820-z} {\bibfield  {journal} {\bibinfo  {journal} {Nature Photonics}\ }\textbf {\bibinfo {volume} {15}},\ \bibinfo {pages} {582} (\bibinfo {year} {2021})}\BibitemShut {NoStop}%
\bibitem [{\citenamefont {Su}\ \emph {et~al.}(2021)\citenamefont {Su}, \citenamefont {Estrecho}, \citenamefont {Biega{\'n}ska}, \citenamefont {Huang}, \citenamefont {Wurdack}, \citenamefont {Pieczarka}, \citenamefont {Truscott}, \citenamefont {Liew}, \citenamefont {Ostrovskaya},\ and\ \citenamefont {Xiong}}]{suDirectMeasurementNonHermitian2021}%
  \BibitemOpen
  \bibfield  {author} {\bibinfo {author} {\bibfnamefont {R.}~\bibnamefont {Su}}, \bibinfo {author} {\bibfnamefont {E.}~\bibnamefont {Estrecho}}, \bibinfo {author} {\bibfnamefont {D.}~\bibnamefont {Biega{\'n}ska}}, \bibinfo {author} {\bibfnamefont {Y.}~\bibnamefont {Huang}}, \bibinfo {author} {\bibfnamefont {M.}~\bibnamefont {Wurdack}}, \bibinfo {author} {\bibfnamefont {M.}~\bibnamefont {Pieczarka}}, \bibinfo {author} {\bibfnamefont {A.~G.}\ \bibnamefont {Truscott}}, \bibinfo {author} {\bibfnamefont {T.~C.~H.}\ \bibnamefont {Liew}}, \bibinfo {author} {\bibfnamefont {E.~A.}\ \bibnamefont {Ostrovskaya}},\ and\ \bibinfo {author} {\bibfnamefont {Q.}~\bibnamefont {Xiong}},\ }\bibfield  {title} {\bibinfo {title} {Direct measurement of a non-{{Hermitian}} topological invariant in a hybrid light-matter system},\ }\bibfield  {journal} {\bibinfo  {journal} {Sci. Adv.}\ }\bibinfo {vol} {7}, \href {https://doi.org/10.1126/sciadv.abj8905} {10.1126/sciadv.abj8905} (\bibinfo {year} {2021})\BibitemShut {NoStop}%
\bibitem [{\citenamefont {Bergholtz}\ \emph {et~al.}(2021)\citenamefont {Bergholtz}, \citenamefont {Budich},\ and\ \citenamefont {Kunst}}]{bergholtzExceptionalTopologyNonHermitian2021}%
  \BibitemOpen
  \bibfield  {author} {\bibinfo {author} {\bibfnamefont {E.~J.}\ \bibnamefont {Bergholtz}}, \bibinfo {author} {\bibfnamefont {J.~C.}\ \bibnamefont {Budich}},\ and\ \bibinfo {author} {\bibfnamefont {F.~K.}\ \bibnamefont {Kunst}},\ }\bibfield  {title} {\bibinfo {title} {Exceptional topology of non-{{Hermitian}} systems},\ }\href {https://doi.org/10.1103/RevModPhys.93.015005} {\bibfield  {journal} {\bibinfo  {journal} {Reviews of Modern Physics}\ }\textbf {\bibinfo {volume} {93}},\ \bibinfo {pages} {015005} (\bibinfo {year} {2021})}\BibitemShut {NoStop}%
\bibitem [{\citenamefont {De~Martini}\ and\ \citenamefont {Jacobovitz}(1988)}]{demartiniAnomalousSpontaneousStimulateddecay1988}%
  \BibitemOpen
  \bibfield  {author} {\bibinfo {author} {\bibfnamefont {F.}~\bibnamefont {De~Martini}}\ and\ \bibinfo {author} {\bibfnamefont {G.~R.}\ \bibnamefont {Jacobovitz}},\ }\bibfield  {title} {\bibinfo {title} {Anomalous spontaneous {\emph{--}} stimulated-decay phase transition and zero-threshold laser action in a microscopic cavity},\ }\href {https://doi.org/10.1103/PhysRevLett.60.1711} {\bibfield  {journal} {\bibinfo  {journal} {Physical Review Letters}\ }\textbf {\bibinfo {volume} {60}},\ \bibinfo {pages} {1711} (\bibinfo {year} {1988})}\BibitemShut {NoStop}%
\bibitem [{\citenamefont {Yokoyama}\ and\ \citenamefont {Brorson}(1989)}]{yokoyamaRateEquationAnalysis1989}%
  \BibitemOpen
  \bibfield  {author} {\bibinfo {author} {\bibfnamefont {H.}~\bibnamefont {Yokoyama}}\ and\ \bibinfo {author} {\bibfnamefont {S.~D.}\ \bibnamefont {Brorson}},\ }\bibfield  {title} {\bibinfo {title} {Rate equation analysis of microcavity lasers},\ }\href {https://doi.org/10.1063/1.343793} {\bibfield  {journal} {\bibinfo  {journal} {Journal of Applied Physics}\ }\textbf {\bibinfo {volume} {66}},\ \bibinfo {pages} {4801} (\bibinfo {year} {1989})}\BibitemShut {NoStop}%
\bibitem [{\citenamefont {Bjork}\ and\ \citenamefont {Yamamoto}(1991)}]{bjorkAnalysisSemiconductorMicrocavity1991}%
  \BibitemOpen
  \bibfield  {author} {\bibinfo {author} {\bibfnamefont {G.}~\bibnamefont {Bjork}}\ and\ \bibinfo {author} {\bibfnamefont {Y.}~\bibnamefont {Yamamoto}},\ }\bibfield  {title} {\bibinfo {title} {Analysis of semiconductor microcavity lasers using rate equations},\ }\href {https://doi.org/10.1109/3.100877} {\bibfield  {journal} {\bibinfo  {journal} {IEEE Journal of Quantum Electronics}\ }\textbf {\bibinfo {volume} {27}},\ \bibinfo {pages} {2386} (\bibinfo {year} {1991})}\BibitemShut {NoStop}%
\bibitem [{\citenamefont {Khurgin}\ and\ \citenamefont {Noginov}(2021)}]{khurginHowPurcellFactor2021}%
  \BibitemOpen
  \bibfield  {author} {\bibinfo {author} {\bibfnamefont {J.~B.}\ \bibnamefont {Khurgin}}\ and\ \bibinfo {author} {\bibfnamefont {M.~A.}\ \bibnamefont {Noginov}},\ }\bibfield  {title} {\bibinfo {title} {How {{Do}} the {{Purcell Factor}}, the {{Q-Factor}}, and the {{Beta Factor Affect}} the {{Laser Threshold}}?},\ }\href {https://doi.org/10.1002/lpor.202000250} {\bibfield  {journal} {\bibinfo  {journal} {Laser \& Photonics Reviews}\ }\textbf {\bibinfo {volume} {15}},\ \bibinfo {pages} {2000250} (\bibinfo {year} {2021})}\BibitemShut {NoStop}%
\bibitem [{\citenamefont {Spillane}\ \emph {et~al.}(2002)\citenamefont {Spillane}, \citenamefont {Kippenberg},\ and\ \citenamefont {Vahala}}]{spillaneUltralowthresholdRamanLaser2002}%
  \BibitemOpen
  \bibfield  {author} {\bibinfo {author} {\bibfnamefont {S.~M.}\ \bibnamefont {Spillane}}, \bibinfo {author} {\bibfnamefont {T.~J.}\ \bibnamefont {Kippenberg}},\ and\ \bibinfo {author} {\bibfnamefont {K.~J.}\ \bibnamefont {Vahala}},\ }\bibfield  {title} {\bibinfo {title} {Ultralow-threshold {{Raman}} laser using a spherical dielectric microcavity},\ }\href {https://doi.org/10.1038/415621a} {\bibfield  {journal} {\bibinfo  {journal} {Nature}\ }\textbf {\bibinfo {volume} {415}},\ \bibinfo {pages} {621} (\bibinfo {year} {2002})}\BibitemShut {NoStop}%
\bibitem [{\citenamefont {Grudinin}\ and\ \citenamefont {Maleki}(2007)}]{grudininUltralowthresholdRamanLasing2007}%
  \BibitemOpen
  \bibfield  {author} {\bibinfo {author} {\bibfnamefont {I.~S.}\ \bibnamefont {Grudinin}}\ and\ \bibinfo {author} {\bibfnamefont {L.}~\bibnamefont {Maleki}},\ }\bibfield  {title} {\bibinfo {title} {Ultralow-threshold {{Raman}} lasing with {{CaF}}{\textsubscript{2}} resonators},\ }\href {https://doi.org/10.1364/OL.32.000166} {\bibfield  {journal} {\bibinfo  {journal} {Optics Letters}\ }\textbf {\bibinfo {volume} {32}},\ \bibinfo {pages} {166} (\bibinfo {year} {2007})}\BibitemShut {NoStop}%
\bibitem [{\citenamefont {Takahashi}\ \emph {et~al.}(2013)\citenamefont {Takahashi}, \citenamefont {Inui}, \citenamefont {Chihara}, \citenamefont {Asano}, \citenamefont {Terawaki},\ and\ \citenamefont {Noda}}]{takahashiMicrometrescaleRamanSilicon2013}%
  \BibitemOpen
  \bibfield  {author} {\bibinfo {author} {\bibfnamefont {Y.}~\bibnamefont {Takahashi}}, \bibinfo {author} {\bibfnamefont {Y.}~\bibnamefont {Inui}}, \bibinfo {author} {\bibfnamefont {M.}~\bibnamefont {Chihara}}, \bibinfo {author} {\bibfnamefont {T.}~\bibnamefont {Asano}}, \bibinfo {author} {\bibfnamefont {R.}~\bibnamefont {Terawaki}},\ and\ \bibinfo {author} {\bibfnamefont {S.}~\bibnamefont {Noda}},\ }\bibfield  {title} {\bibinfo {title} {A micrometre-scale {{Raman}} silicon laser with a microwatt threshold},\ }\href {https://doi.org/10.1038/nature12237} {\bibfield  {journal} {\bibinfo  {journal} {Nature}\ }\textbf {\bibinfo {volume} {498}},\ \bibinfo {pages} {470} (\bibinfo {year} {2013})}\BibitemShut {NoStop}%
\bibitem [{\citenamefont {Tyumenev}\ \emph {et~al.}(2022)\citenamefont {Tyumenev}, \citenamefont {Hammer}, \citenamefont {Joly}, \citenamefont {Russell},\ and\ \citenamefont {Novoa}}]{tyumenevTunableStatepreservingFrequency2022}%
  \BibitemOpen
  \bibfield  {author} {\bibinfo {author} {\bibfnamefont {R.}~\bibnamefont {Tyumenev}}, \bibinfo {author} {\bibfnamefont {J.}~\bibnamefont {Hammer}}, \bibinfo {author} {\bibfnamefont {N.~Y.}\ \bibnamefont {Joly}}, \bibinfo {author} {\bibfnamefont {P.~{\relax St}.~J.}\ \bibnamefont {Russell}},\ and\ \bibinfo {author} {\bibfnamefont {D.}~\bibnamefont {Novoa}},\ }\bibfield  {title} {\bibinfo {title} {Tunable and state-preserving frequency conversion of single photons in hydrogen},\ }\href {https://doi.org/10.1126/science.abn1434} {\bibfield  {journal} {\bibinfo  {journal} {Science}\ }\textbf {\bibinfo {volume} {376}},\ \bibinfo {pages} {621} (\bibinfo {year} {2022})}\BibitemShut {NoStop}%
\bibitem [{\citenamefont {Feng}\ \emph {et~al.}(2014)\citenamefont {Feng}, \citenamefont {Wong}, \citenamefont {Ma}, \citenamefont {Wang},\ and\ \citenamefont {Zhang}}]{fengSinglemodeLaserParitytime2014}%
  \BibitemOpen
  \bibfield  {author} {\bibinfo {author} {\bibfnamefont {L.}~\bibnamefont {Feng}}, \bibinfo {author} {\bibfnamefont {Z.~J.}\ \bibnamefont {Wong}}, \bibinfo {author} {\bibfnamefont {R.-M.}\ \bibnamefont {Ma}}, \bibinfo {author} {\bibfnamefont {Y.}~\bibnamefont {Wang}},\ and\ \bibinfo {author} {\bibfnamefont {X.}~\bibnamefont {Zhang}},\ }\bibfield  {title} {\bibinfo {title} {Single-mode laser by parity-time symmetry breaking},\ }\href {https://doi.org/10.1126/science.1258479} {\bibfield  {journal} {\bibinfo  {journal} {Science}\ }\textbf {\bibinfo {volume} {346}},\ \bibinfo {pages} {972} (\bibinfo {year} {2014})}\BibitemShut {NoStop}%
\bibitem [{\citenamefont {Guo}\ \emph {et~al.}(2009)\citenamefont {Guo}, \citenamefont {Salamo}, \citenamefont {Duchesne}, \citenamefont {Morandotti}, \citenamefont {{Volatier-Ravat}}, \citenamefont {Aimez}, \citenamefont {Siviloglou},\ and\ \citenamefont {Christodoulides}}]{guoObservation$mathcalPmathcalT$SymmetryBreaking2009}%
  \BibitemOpen
  \bibfield  {author} {\bibinfo {author} {\bibfnamefont {A.}~\bibnamefont {Guo}}, \bibinfo {author} {\bibfnamefont {G.~J.}\ \bibnamefont {Salamo}}, \bibinfo {author} {\bibfnamefont {D.}~\bibnamefont {Duchesne}}, \bibinfo {author} {\bibfnamefont {R.}~\bibnamefont {Morandotti}}, \bibinfo {author} {\bibfnamefont {M.}~\bibnamefont {{Volatier-Ravat}}}, \bibinfo {author} {\bibfnamefont {V.}~\bibnamefont {Aimez}}, \bibinfo {author} {\bibfnamefont {G.~A.}\ \bibnamefont {Siviloglou}},\ and\ \bibinfo {author} {\bibfnamefont {D.~N.}\ \bibnamefont {Christodoulides}},\ }\bibfield  {title} {\bibinfo {title} {Observation of PT-{{Symmetry Breaking}} in {{Complex Optical Potentials}}},\ }\href {https://doi.org/10.1103/PhysRevLett.103.093902} {\bibfield  {journal} {\bibinfo  {journal} {Physical Review Letters}\ }\textbf {\bibinfo {volume} {103}},\ \bibinfo {pages} {093902} (\bibinfo {year} {2009})}\BibitemShut
  {NoStop}%
\bibitem [{\citenamefont {Peng}\ \emph {et~al.}(2016)\citenamefont {Peng}, \citenamefont {{\"O}zdemir}, \citenamefont {Liertzer}, \citenamefont {Chen}, \citenamefont {Kramer}, \citenamefont {Y{\i}lmaz}, \citenamefont {Wiersig}, \citenamefont {Rotter},\ and\ \citenamefont {Yang}}]{pengChiralModesDirectional2016}%
  \BibitemOpen
  \bibfield  {author} {\bibinfo {author} {\bibfnamefont {B.}~\bibnamefont {Peng}}, \bibinfo {author} {\bibfnamefont {{\c S}.~K.}\ \bibnamefont {{\"O}zdemir}}, \bibinfo {author} {\bibfnamefont {M.}~\bibnamefont {Liertzer}}, \bibinfo {author} {\bibfnamefont {W.}~\bibnamefont {Chen}}, \bibinfo {author} {\bibfnamefont {J.}~\bibnamefont {Kramer}}, \bibinfo {author} {\bibfnamefont {H.}~\bibnamefont {Y{\i}lmaz}}, \bibinfo {author} {\bibfnamefont {J.}~\bibnamefont {Wiersig}}, \bibinfo {author} {\bibfnamefont {S.}~\bibnamefont {Rotter}},\ and\ \bibinfo {author} {\bibfnamefont {L.}~\bibnamefont {Yang}},\ }\bibfield  {title} {\bibinfo {title} {Chiral modes and directional lasing at exceptional points},\ }\href {https://doi.org/10.1073/pnas.1603318113} {\bibfield  {journal} {\bibinfo  {journal} {Proceedings of the National Academy of Sciences}\ }\textbf {\bibinfo {volume} {113}},\ \bibinfo {pages} {6845} (\bibinfo {year} {2016})}\BibitemShut {NoStop}%
\bibitem [{\citenamefont {Kepesidis}\ \emph {et~al.}(2016)\citenamefont {Kepesidis}, \citenamefont {Milburn}, \citenamefont {Huber}, \citenamefont {Makris}, \citenamefont {Rotter},\ and\ \citenamefont {Rabl}}]{kepesidisPTsymmetryBreakingSteady2016}%
  \BibitemOpen
  \bibfield  {author} {\bibinfo {author} {\bibfnamefont {K.~V.}\ \bibnamefont {Kepesidis}}, \bibinfo {author} {\bibfnamefont {T.~J.}\ \bibnamefont {Milburn}}, \bibinfo {author} {\bibfnamefont {J.}~\bibnamefont {Huber}}, \bibinfo {author} {\bibfnamefont {K.~G.}\ \bibnamefont {Makris}}, \bibinfo {author} {\bibfnamefont {S.}~\bibnamefont {Rotter}},\ and\ \bibinfo {author} {\bibfnamefont {P.}~\bibnamefont {Rabl}},\ }\bibfield  {title} {\bibinfo {title} {{{PT-symmetry}} breaking in the steady state of microscopic gain--loss systems},\ }\href {https://doi.org/10.1088/1367-2630/18/9/095003} {\bibfield  {journal} {\bibinfo  {journal} {New Journal of Physics}\ }\textbf {\bibinfo {volume} {18}},\ \bibinfo {pages} {095003} (\bibinfo {year} {2016})}\BibitemShut {NoStop}%
\bibitem [{\citenamefont {Zhang}\ \emph {et~al.}(2017)\citenamefont {Zhang}, \citenamefont {Agarwal}, \citenamefont {Schleich},\ and\ \citenamefont {Scully}}]{zhangHiddenPTSymmetry2017}%
  \BibitemOpen
  \bibfield  {author} {\bibinfo {author} {\bibfnamefont {L.}~\bibnamefont {Zhang}}, \bibinfo {author} {\bibfnamefont {G.~S.}\ \bibnamefont {Agarwal}}, \bibinfo {author} {\bibfnamefont {W.~P.}\ \bibnamefont {Schleich}},\ and\ \bibinfo {author} {\bibfnamefont {M.~O.}\ \bibnamefont {Scully}},\ }\bibfield  {title} {\bibinfo {title} {Hidden {{PT}} symmetry and quantization of a coupled-oscillator model of quantum amplification by superradiant emission of radiation},\ }\href {https://doi.org/10.1103/PhysRevA.96.013827} {\bibfield  {journal} {\bibinfo  {journal} {Physical Review A}\ }\textbf {\bibinfo {volume} {96}},\ \bibinfo {pages} {013827} (\bibinfo {year} {2017})}\BibitemShut {NoStop}%
\bibitem [{\citenamefont {Feng}\ \emph {et~al.}(2017)\citenamefont {Feng}, \citenamefont {{El-Ganainy}},\ and\ \citenamefont {Ge}}]{fengNonHermitianPhotonicsBased2017}%
  \BibitemOpen
  \bibfield  {author} {\bibinfo {author} {\bibfnamefont {L.}~\bibnamefont {Feng}}, \bibinfo {author} {\bibfnamefont {R.}~\bibnamefont {{El-Ganainy}}},\ and\ \bibinfo {author} {\bibfnamefont {L.}~\bibnamefont {Ge}},\ }\bibfield  {title} {\bibinfo {title} {Non-{{Hermitian}} photonics based on parity--time symmetry},\ }\href {https://doi.org/10.1038/s41566-017-0031-1} {\bibfield  {journal} {\bibinfo  {journal} {Nature Photonics}\ }\textbf {\bibinfo {volume} {11}},\ \bibinfo {pages} {752} (\bibinfo {year} {2017})}\BibitemShut {NoStop}%
\bibitem [{\citenamefont {{El-Ganainy}}\ \emph {et~al.}(2018)\citenamefont {{El-Ganainy}}, \citenamefont {Makris}, \citenamefont {Khajavikhan}, \citenamefont {Musslimani}, \citenamefont {Rotter},\ and\ \citenamefont {Christodoulides}}]{el-ganainyNonHermitianPhysicsPT2018}%
  \BibitemOpen
  \bibfield  {author} {\bibinfo {author} {\bibfnamefont {R.}~\bibnamefont {{El-Ganainy}}}, \bibinfo {author} {\bibfnamefont {K.~G.}\ \bibnamefont {Makris}}, \bibinfo {author} {\bibfnamefont {M.}~\bibnamefont {Khajavikhan}}, \bibinfo {author} {\bibfnamefont {Z.~H.}\ \bibnamefont {Musslimani}}, \bibinfo {author} {\bibfnamefont {S.}~\bibnamefont {Rotter}},\ and\ \bibinfo {author} {\bibfnamefont {D.~N.}\ \bibnamefont {Christodoulides}},\ }\bibfield  {title} {\bibinfo {title} {Non-{{Hermitian}} physics and {{PT}} symmetry},\ }\href {https://doi.org/10.1038/nphys4323} {\bibfield  {journal} {\bibinfo  {journal} {Nature Physics}\ }\textbf {\bibinfo {volume} {14}},\ \bibinfo {pages} {11} (\bibinfo {year} {2018})}\BibitemShut {NoStop}%
\bibitem [{\citenamefont {Li}\ \emph {et~al.}(2022)\citenamefont {Li}, \citenamefont {Ma}, \citenamefont {Hatzopoulos}, \citenamefont {Savvidis}, \citenamefont {Schumacher},\ and\ \citenamefont {Gao}}]{liSwitchingMicrocavityPolariton2022}%
  \BibitemOpen
  \bibfield  {author} {\bibinfo {author} {\bibfnamefont {Y.}~\bibnamefont {Li}}, \bibinfo {author} {\bibfnamefont {X.}~\bibnamefont {Ma}}, \bibinfo {author} {\bibfnamefont {Z.}~\bibnamefont {Hatzopoulos}}, \bibinfo {author} {\bibfnamefont {P.~G.}\ \bibnamefont {Savvidis}}, \bibinfo {author} {\bibfnamefont {S.}~\bibnamefont {Schumacher}},\ and\ \bibinfo {author} {\bibfnamefont {T.}~\bibnamefont {Gao}},\ }\bibfield  {title} {\bibinfo {title} {Switching {{Off}} a {{Microcavity Polariton Condensate}} near the {{Exceptional Point}}},\ }\href {https://doi.org/10.1021/acsphotonics.2c00288} {\bibfield  {journal} {\bibinfo  {journal} {ACS Photonics}\ }\textbf {\bibinfo {volume} {9}},\ \bibinfo {pages} {2079} (\bibinfo {year} {2022})}\BibitemShut {NoStop}%
\bibitem [{\citenamefont {R{\"u}ter}\ \emph {et~al.}(2010)\citenamefont {R{\"u}ter}, \citenamefont {Makris}, \citenamefont {{El-Ganainy}}, \citenamefont {Christodoulides}, \citenamefont {Segev},\ and\ \citenamefont {Kip}}]{ruterObservationParityTime2010}%
  \BibitemOpen
  \bibfield  {author} {\bibinfo {author} {\bibfnamefont {C.~E.}\ \bibnamefont {R{\"u}ter}}, \bibinfo {author} {\bibfnamefont {K.~G.}\ \bibnamefont {Makris}}, \bibinfo {author} {\bibfnamefont {R.}~\bibnamefont {{El-Ganainy}}}, \bibinfo {author} {\bibfnamefont {D.~N.}\ \bibnamefont {Christodoulides}}, \bibinfo {author} {\bibfnamefont {M.}~\bibnamefont {Segev}},\ and\ \bibinfo {author} {\bibfnamefont {D.}~\bibnamefont {Kip}},\ }\bibfield  {title} {\bibinfo {title} {Observation of parity--time symmetry in optics},\ }\href {https://doi.org/10.1038/nphys1515} {\bibfield  {journal} {\bibinfo  {journal} {Nature Physics}\ }\textbf {\bibinfo {volume} {6}},\ \bibinfo {pages} {192} (\bibinfo {year} {2010})}\BibitemShut {NoStop}%
\bibitem [{\citenamefont {Hodaei}\ \emph {et~al.}(2014)\citenamefont {Hodaei}, \citenamefont {Miri}, \citenamefont {Heinrich}, \citenamefont {Christodoulides},\ and\ \citenamefont {Khajavikhan}}]{hodaeiParitytimesymmetricMicroringLasers2014}%
  \BibitemOpen
  \bibfield  {author} {\bibinfo {author} {\bibfnamefont {H.}~\bibnamefont {Hodaei}}, \bibinfo {author} {\bibfnamefont {M.-A.}\ \bibnamefont {Miri}}, \bibinfo {author} {\bibfnamefont {M.}~\bibnamefont {Heinrich}}, \bibinfo {author} {\bibfnamefont {D.~N.}\ \bibnamefont {Christodoulides}},\ and\ \bibinfo {author} {\bibfnamefont {M.}~\bibnamefont {Khajavikhan}},\ }\bibfield  {title} {\bibinfo {title} {Parity-time--symmetric microring lasers},\ }\href {https://doi.org/10.1126/science.1258480} {\bibfield  {journal} {\bibinfo  {journal} {Science}\ }\textbf {\bibinfo {volume} {346}},\ \bibinfo {pages} {975} (\bibinfo {year} {2014})}\BibitemShut {NoStop}%
\bibitem [{\citenamefont {Wu}\ \emph {et~al.}(2015)\citenamefont {Wu}, \citenamefont {Buckley}, \citenamefont {Schaibley}, \citenamefont {Feng}, \citenamefont {Yan}, \citenamefont {Mandrus}, \citenamefont {Hatami}, \citenamefont {Yao}, \citenamefont {Vu{\v c}kovi{\'c}}, \citenamefont {Majumdar},\ and\ \citenamefont {Xu}}]{wuMonolayerSemiconductorNanocavity2015}%
  \BibitemOpen
  \bibfield  {author} {\bibinfo {author} {\bibfnamefont {S.}~\bibnamefont {Wu}}, \bibinfo {author} {\bibfnamefont {S.}~\bibnamefont {Buckley}}, \bibinfo {author} {\bibfnamefont {J.~R.}\ \bibnamefont {Schaibley}}, \bibinfo {author} {\bibfnamefont {L.}~\bibnamefont {Feng}}, \bibinfo {author} {\bibfnamefont {J.}~\bibnamefont {Yan}}, \bibinfo {author} {\bibfnamefont {D.~G.}\ \bibnamefont {Mandrus}}, \bibinfo {author} {\bibfnamefont {F.}~\bibnamefont {Hatami}}, \bibinfo {author} {\bibfnamefont {W.}~\bibnamefont {Yao}}, \bibinfo {author} {\bibfnamefont {J.}~\bibnamefont {Vu{\v c}kovi{\'c}}}, \bibinfo {author} {\bibfnamefont {A.}~\bibnamefont {Majumdar}},\ and\ \bibinfo {author} {\bibfnamefont {X.}~\bibnamefont {Xu}},\ }\bibfield  {title} {\bibinfo {title} {Monolayer semiconductor nanocavity lasers with ultralow thresholds},\ }\href {https://doi.org/10.1038/nature14290} {\bibfield  {journal} {\bibinfo  {journal} {Nature}\ }\textbf {\bibinfo {volume} {520}},\ \bibinfo {pages} {69} (\bibinfo {year}
  {2015})}\BibitemShut {NoStop}%
\bibitem [{\citenamefont {Ye}\ \emph {et~al.}(2015)\citenamefont {Ye}, \citenamefont {Wong}, \citenamefont {Lu}, \citenamefont {Ni}, \citenamefont {Zhu}, \citenamefont {Chen}, \citenamefont {Wang},\ and\ \citenamefont {Zhang}}]{yeMonolayerExcitonicLaser2015}%
  \BibitemOpen
  \bibfield  {author} {\bibinfo {author} {\bibfnamefont {Y.}~\bibnamefont {Ye}}, \bibinfo {author} {\bibfnamefont {Z.~J.}\ \bibnamefont {Wong}}, \bibinfo {author} {\bibfnamefont {X.}~\bibnamefont {Lu}}, \bibinfo {author} {\bibfnamefont {X.}~\bibnamefont {Ni}}, \bibinfo {author} {\bibfnamefont {H.}~\bibnamefont {Zhu}}, \bibinfo {author} {\bibfnamefont {X.}~\bibnamefont {Chen}}, \bibinfo {author} {\bibfnamefont {Y.}~\bibnamefont {Wang}},\ and\ \bibinfo {author} {\bibfnamefont {X.}~\bibnamefont {Zhang}},\ }\bibfield  {title} {\bibinfo {title} {Monolayer excitonic laser},\ }\href {https://doi.org/10.1038/nphoton.2015.197} {\bibfield  {journal} {\bibinfo  {journal} {Nature Photonics}\ }\textbf {\bibinfo {volume} {9}},\ \bibinfo {pages} {733} (\bibinfo {year} {2015})}\BibitemShut {NoStop}%
\bibitem [{\citenamefont {Shang}\ \emph {et~al.}(2017)\citenamefont {Shang}, \citenamefont {Cong}, \citenamefont {Wang}, \citenamefont {Peimyoo}, \citenamefont {Wu}, \citenamefont {Zou}, \citenamefont {Chen}, \citenamefont {Chin}, \citenamefont {Wang}, \citenamefont {Soci}, \citenamefont {Huang},\ and\ \citenamefont {Yu}}]{shangRoomtemperature2DSemiconductor2017}%
  \BibitemOpen
  \bibfield  {author} {\bibinfo {author} {\bibfnamefont {J.}~\bibnamefont {Shang}}, \bibinfo {author} {\bibfnamefont {C.}~\bibnamefont {Cong}}, \bibinfo {author} {\bibfnamefont {Z.}~\bibnamefont {Wang}}, \bibinfo {author} {\bibfnamefont {N.}~\bibnamefont {Peimyoo}}, \bibinfo {author} {\bibfnamefont {L.}~\bibnamefont {Wu}}, \bibinfo {author} {\bibfnamefont {C.}~\bibnamefont {Zou}}, \bibinfo {author} {\bibfnamefont {Y.}~\bibnamefont {Chen}}, \bibinfo {author} {\bibfnamefont {X.~Y.}\ \bibnamefont {Chin}}, \bibinfo {author} {\bibfnamefont {J.}~\bibnamefont {Wang}}, \bibinfo {author} {\bibfnamefont {C.}~\bibnamefont {Soci}}, \bibinfo {author} {\bibfnamefont {W.}~\bibnamefont {Huang}},\ and\ \bibinfo {author} {\bibfnamefont {T.}~\bibnamefont {Yu}},\ }\bibfield  {title} {\bibinfo {title} {Room-temperature {{2D}} semiconductor activated vertical-cavity surface-emitting lasers},\ }\href {https://doi.org/10.1038/s41467-017-00743-w} {\bibfield  {journal} {\bibinfo  {journal} {Nature Communications}\ }\textbf {\bibinfo
  {volume} {8}},\ \bibinfo {pages} {543} (\bibinfo {year} {2017})}\BibitemShut {NoStop}%
\bibitem [{\citenamefont {Liu}\ \emph {et~al.}(2015)\citenamefont {Liu}, \citenamefont {Galfsky}, \citenamefont {Sun}, \citenamefont {Xia}, \citenamefont {Lin}, \citenamefont {Lee}, \citenamefont {{K{\'e}na-Cohen}},\ and\ \citenamefont {Menon}}]{liuStrongLightMatter2015}%
  \BibitemOpen
  \bibfield  {author} {\bibinfo {author} {\bibfnamefont {X.}~\bibnamefont {Liu}}, \bibinfo {author} {\bibfnamefont {T.}~\bibnamefont {Galfsky}}, \bibinfo {author} {\bibfnamefont {Z.}~\bibnamefont {Sun}}, \bibinfo {author} {\bibfnamefont {F.}~\bibnamefont {Xia}}, \bibinfo {author} {\bibfnamefont {E.-c.}\ \bibnamefont {Lin}}, \bibinfo {author} {\bibfnamefont {Y.-H.}\ \bibnamefont {Lee}}, \bibinfo {author} {\bibfnamefont {S.}~\bibnamefont {{K{\'e}na-Cohen}}},\ and\ \bibinfo {author} {\bibfnamefont {V.~M.}\ \bibnamefont {Menon}},\ }\bibfield  {title} {\bibinfo {title} {Strong light--matter coupling in two-dimensional atomic crystals},\ }\href {https://doi.org/10.1038/nphoton.2014.304} {\bibfield  {journal} {\bibinfo  {journal} {Nature Photonics}\ }\textbf {\bibinfo {volume} {9}},\ \bibinfo {pages} {30} (\bibinfo {year} {2015})}\BibitemShut {NoStop}%
\bibitem [{\citenamefont {Dhara}\ \emph {et~al.}(2018)\citenamefont {Dhara}, \citenamefont {Chakraborty}, \citenamefont {Goodfellow}, \citenamefont {Qiu}, \citenamefont {O'Loughlin}, \citenamefont {Wicks}, \citenamefont {Bhattacharjee},\ and\ \citenamefont {Vamivakas}}]{dharaAnomalousDispersionMicrocavity2018}%
  \BibitemOpen
  \bibfield  {author} {\bibinfo {author} {\bibfnamefont {S.}~\bibnamefont {Dhara}}, \bibinfo {author} {\bibfnamefont {C.}~\bibnamefont {Chakraborty}}, \bibinfo {author} {\bibfnamefont {K.~M.}\ \bibnamefont {Goodfellow}}, \bibinfo {author} {\bibfnamefont {L.}~\bibnamefont {Qiu}}, \bibinfo {author} {\bibfnamefont {T.~A.}\ \bibnamefont {O'Loughlin}}, \bibinfo {author} {\bibfnamefont {G.~W.}\ \bibnamefont {Wicks}}, \bibinfo {author} {\bibfnamefont {S.}~\bibnamefont {Bhattacharjee}},\ and\ \bibinfo {author} {\bibfnamefont {A.~N.}\ \bibnamefont {Vamivakas}},\ }\bibfield  {title} {\bibinfo {title} {Anomalous dispersion of microcavity trion-polaritons},\ }\href {https://doi.org/10.1038/nphys4303} {\bibfield  {journal} {\bibinfo  {journal} {Nature Physics}\ }\textbf {\bibinfo {volume} {14}},\ \bibinfo {pages} {130} (\bibinfo {year} {2018})}\BibitemShut {NoStop}%
\bibitem [{\citenamefont {Aslan}\ \emph {et~al.}(2016)\citenamefont {Aslan}, \citenamefont {Chenet}, \citenamefont {{van der Zande}}, \citenamefont {Hone},\ and\ \citenamefont {Heinz}}]{aslanLinearlyPolarizedExcitons2016}%
  \BibitemOpen
  \bibfield  {author} {\bibinfo {author} {\bibfnamefont {O.~B.}\ \bibnamefont {Aslan}}, \bibinfo {author} {\bibfnamefont {D.~A.}\ \bibnamefont {Chenet}}, \bibinfo {author} {\bibfnamefont {A.~M.}\ \bibnamefont {{van der Zande}}}, \bibinfo {author} {\bibfnamefont {J.~C.}\ \bibnamefont {Hone}},\ and\ \bibinfo {author} {\bibfnamefont {T.~F.}\ \bibnamefont {Heinz}},\ }\bibfield  {title} {\bibinfo {title} {Linearly {{Polarized Excitons}} in {{Single-}} and {{Few-Layer ReS}} {\textsubscript{2}} {{Crystals}}},\ }\href {https://doi.org/10.1021/acsphotonics.5b00486} {\bibfield  {journal} {\bibinfo  {journal} {ACS Photonics}\ }\textbf {\bibinfo {volume} {3}},\ \bibinfo {pages} {96} (\bibinfo {year} {2016})}\BibitemShut {NoStop}%
\bibitem [{\citenamefont {Dhara}\ \emph {et~al.}(2020)\citenamefont {Dhara}, \citenamefont {Chakrabarty}, \citenamefont {Das}, \citenamefont {Pattanayak}, \citenamefont {Paul}, \citenamefont {Mukherjee},\ and\ \citenamefont {Dhara}}]{dharaAdditionalExcitonicFeatures2020}%
  \BibitemOpen
  \bibfield  {author} {\bibinfo {author} {\bibfnamefont {A.}~\bibnamefont {Dhara}}, \bibinfo {author} {\bibfnamefont {D.}~\bibnamefont {Chakrabarty}}, \bibinfo {author} {\bibfnamefont {P.}~\bibnamefont {Das}}, \bibinfo {author} {\bibfnamefont {A.~K.}\ \bibnamefont {Pattanayak}}, \bibinfo {author} {\bibfnamefont {S.}~\bibnamefont {Paul}}, \bibinfo {author} {\bibfnamefont {S.}~\bibnamefont {Mukherjee}},\ and\ \bibinfo {author} {\bibfnamefont {S.}~\bibnamefont {Dhara}},\ }\bibfield  {title} {\bibinfo {title} {Additional excitonic features and momentum-dark states in {{ReS2}}},\ }\href {https://doi.org/10.1103/PhysRevB.102.161404} {\bibfield  {journal} {\bibinfo  {journal} {Physical Review B}\ }\textbf {\bibinfo {volume} {102}},\ \bibinfo {pages} {161404(R)} (\bibinfo {year} {2020})}\BibitemShut {NoStop}%
\bibitem [{\citenamefont {Chakrabarty}\ \emph {et~al.}(2021)\citenamefont {Chakrabarty}, \citenamefont {Dhara}, \citenamefont {Ghosh}, \citenamefont {Pattanayak}, \citenamefont {Mukherjee}, \citenamefont {Chaudhuri},\ and\ \citenamefont {Dhara}}]{chakrabartyInterfacialAnisotropicExcitonpolariton2021}%
  \BibitemOpen
  \bibfield  {author} {\bibinfo {author} {\bibfnamefont {D.}~\bibnamefont {Chakrabarty}}, \bibinfo {author} {\bibfnamefont {A.}~\bibnamefont {Dhara}}, \bibinfo {author} {\bibfnamefont {K.}~\bibnamefont {Ghosh}}, \bibinfo {author} {\bibfnamefont {A.~K.}\ \bibnamefont {Pattanayak}}, \bibinfo {author} {\bibfnamefont {S.}~\bibnamefont {Mukherjee}}, \bibinfo {author} {\bibfnamefont {A.~R.}\ \bibnamefont {Chaudhuri}},\ and\ \bibinfo {author} {\bibfnamefont {S.}~\bibnamefont {Dhara}},\ }\bibfield  {title} {\bibinfo {title} {Interfacial anisotropic exciton-polariton manifolds in {{ReS}}{\textsubscript{2}}},\ }\href {https://doi.org/10.1364/OPTICA.435647} {\bibfield  {journal} {\bibinfo  {journal} {Optica}\ }\textbf {\bibinfo {volume} {8}},\ \bibinfo {pages} {1488} (\bibinfo {year} {2021})}\BibitemShut {NoStop}%
\bibitem [{\citenamefont {Peng}\ \emph {et~al.}(2014)\citenamefont {Peng}, \citenamefont {{\"O}zdemir}, \citenamefont {Lei}, \citenamefont {Monifi}, \citenamefont {Gianfreda}, \citenamefont {Long}, \citenamefont {Fan}, \citenamefont {Nori}, \citenamefont {Bender},\ and\ \citenamefont {Yang}}]{pengParityTimeSymmetric2014}%
  \BibitemOpen
  \bibfield  {author} {\bibinfo {author} {\bibfnamefont {B.}~\bibnamefont {Peng}}, \bibinfo {author} {\bibfnamefont {{\c S}.~K.}\ \bibnamefont {{\"O}zdemir}}, \bibinfo {author} {\bibfnamefont {F.}~\bibnamefont {Lei}}, \bibinfo {author} {\bibfnamefont {F.}~\bibnamefont {Monifi}}, \bibinfo {author} {\bibfnamefont {M.}~\bibnamefont {Gianfreda}}, \bibinfo {author} {\bibfnamefont {G.~L.}\ \bibnamefont {Long}}, \bibinfo {author} {\bibfnamefont {S.}~\bibnamefont {Fan}}, \bibinfo {author} {\bibfnamefont {F.}~\bibnamefont {Nori}}, \bibinfo {author} {\bibfnamefont {C.~M.}\ \bibnamefont {Bender}},\ and\ \bibinfo {author} {\bibfnamefont {L.}~\bibnamefont {Yang}},\ }\bibfield  {title} {\bibinfo {title} {Parity--time-symmetric whispering-gallery microcavities},\ }\href {https://doi.org/10.1038/nphys2927} {\bibfield  {journal} {\bibinfo  {journal} {Nature Physics}\ }\textbf {\bibinfo {volume} {10}},\ \bibinfo {pages} {394} (\bibinfo {year} {2014})}\BibitemShut {NoStop}%
\bibitem [{\citenamefont {Chang}\ \emph {et~al.}(2014)\citenamefont {Chang}, \citenamefont {Jiang}, \citenamefont {Hua}, \citenamefont {Yang}, \citenamefont {Wen}, \citenamefont {Jiang}, \citenamefont {Li}, \citenamefont {Wang},\ and\ \citenamefont {Xiao}}]{changParityTimeSymmetry2014}%
  \BibitemOpen
  \bibfield  {author} {\bibinfo {author} {\bibfnamefont {L.}~\bibnamefont {Chang}}, \bibinfo {author} {\bibfnamefont {X.}~\bibnamefont {Jiang}}, \bibinfo {author} {\bibfnamefont {S.}~\bibnamefont {Hua}}, \bibinfo {author} {\bibfnamefont {C.}~\bibnamefont {Yang}}, \bibinfo {author} {\bibfnamefont {J.}~\bibnamefont {Wen}}, \bibinfo {author} {\bibfnamefont {L.}~\bibnamefont {Jiang}}, \bibinfo {author} {\bibfnamefont {G.}~\bibnamefont {Li}}, \bibinfo {author} {\bibfnamefont {G.}~\bibnamefont {Wang}},\ and\ \bibinfo {author} {\bibfnamefont {M.}~\bibnamefont {Xiao}},\ }\bibfield  {title} {\bibinfo {title} {Parity--time symmetry and variable optical isolation in active--passive-coupled microresonators},\ }\href {https://doi.org/10.1038/nphoton.2014.133} {\bibfield  {journal} {\bibinfo  {journal} {Nature Photonics}\ }\textbf {\bibinfo {volume} {8}},\ \bibinfo {pages} {524} (\bibinfo {year} {2014})}\BibitemShut {NoStop}%
\bibitem [{\citenamefont {Gao}\ \emph {et~al.}(2018{\natexlab{b}})\citenamefont {Gao}, \citenamefont {Li}, \citenamefont {Estrecho}, \citenamefont {Liew}, \citenamefont {{Comber-Todd}}, \citenamefont {Nalitov}, \citenamefont {Steger}, \citenamefont {West}, \citenamefont {Pfeiffer}, \citenamefont {Snoke}, \citenamefont {Kavokin}, \citenamefont {Truscott},\ and\ \citenamefont {Ostrovskaya}}]{gaoChiralModesExceptional2018}%
  \BibitemOpen
  \bibfield  {author} {\bibinfo {author} {\bibfnamefont {T.}~\bibnamefont {Gao}}, \bibinfo {author} {\bibfnamefont {G.}~\bibnamefont {Li}}, \bibinfo {author} {\bibfnamefont {E.}~\bibnamefont {Estrecho}}, \bibinfo {author} {\bibfnamefont {T.~C.~H.}\ \bibnamefont {Liew}}, \bibinfo {author} {\bibfnamefont {D.}~\bibnamefont {{Comber-Todd}}}, \bibinfo {author} {\bibfnamefont {A.}~\bibnamefont {Nalitov}}, \bibinfo {author} {\bibfnamefont {M.}~\bibnamefont {Steger}}, \bibinfo {author} {\bibfnamefont {K.}~\bibnamefont {West}}, \bibinfo {author} {\bibfnamefont {L.}~\bibnamefont {Pfeiffer}}, \bibinfo {author} {\bibfnamefont {D.~W.}\ \bibnamefont {Snoke}}, \bibinfo {author} {\bibfnamefont {A.~V.}\ \bibnamefont {Kavokin}}, \bibinfo {author} {\bibfnamefont {A.~G.}\ \bibnamefont {Truscott}},\ and\ \bibinfo {author} {\bibfnamefont {E.~A.}\ \bibnamefont {Ostrovskaya}},\ }\bibfield  {title} {\bibinfo {title} {Chiral {{Modes}} at {{Exceptional Points}} in {{Exciton-Polariton Quantum Fluids}}},\ }\href
  {https://doi.org/10.1103/PhysRevLett.120.065301} {\bibfield  {journal} {\bibinfo  {journal} {Physical Review Letters}\ }\textbf {\bibinfo {volume} {120}},\ \bibinfo {pages} {065301} (\bibinfo {year} {2018}{\natexlab{b}})}\BibitemShut {NoStop}%
\bibitem [{\citenamefont {Zhang}\ \emph {et~al.}(2018)\citenamefont {Zhang}, \citenamefont {Peng}, \citenamefont {{\"O}zdemir}, \citenamefont {Pichler}, \citenamefont {Krimer}, \citenamefont {Zhao}, \citenamefont {Nori}, \citenamefont {Liu}, \citenamefont {Rotter},\ and\ \citenamefont {Yang}}]{zhangPhononLaserOperating2018}%
  \BibitemOpen
  \bibfield  {author} {\bibinfo {author} {\bibfnamefont {J.}~\bibnamefont {Zhang}}, \bibinfo {author} {\bibfnamefont {B.}~\bibnamefont {Peng}}, \bibinfo {author} {\bibfnamefont {{\c S}.~K.}\ \bibnamefont {{\"O}zdemir}}, \bibinfo {author} {\bibfnamefont {K.}~\bibnamefont {Pichler}}, \bibinfo {author} {\bibfnamefont {D.~O.}\ \bibnamefont {Krimer}}, \bibinfo {author} {\bibfnamefont {G.}~\bibnamefont {Zhao}}, \bibinfo {author} {\bibfnamefont {F.}~\bibnamefont {Nori}}, \bibinfo {author} {\bibfnamefont {Y.-x.}\ \bibnamefont {Liu}}, \bibinfo {author} {\bibfnamefont {S.}~\bibnamefont {Rotter}},\ and\ \bibinfo {author} {\bibfnamefont {L.}~\bibnamefont {Yang}},\ }\bibfield  {title} {\bibinfo {title} {A phonon laser operating at an exceptional point},\ }\href {https://doi.org/10.1038/s41566-018-0213-5} {\bibfield  {journal} {\bibinfo  {journal} {Nature Photonics}\ }\textbf {\bibinfo {volume} {12}},\ \bibinfo {pages} {479} (\bibinfo {year} {2018})}\BibitemShut {NoStop}%
\bibitem [{\citenamefont {Kremer}\ \emph {et~al.}(2019)\citenamefont {Kremer}, \citenamefont {Biesenthal}, \citenamefont {Maczewsky}, \citenamefont {Heinrich}, \citenamefont {Thomale},\ and\ \citenamefont {Szameit}}]{kremerDemonstrationTwodimensionalPTsymmetric2019}%
  \BibitemOpen
  \bibfield  {author} {\bibinfo {author} {\bibfnamefont {M.}~\bibnamefont {Kremer}}, \bibinfo {author} {\bibfnamefont {T.}~\bibnamefont {Biesenthal}}, \bibinfo {author} {\bibfnamefont {L.~J.}\ \bibnamefont {Maczewsky}}, \bibinfo {author} {\bibfnamefont {M.}~\bibnamefont {Heinrich}}, \bibinfo {author} {\bibfnamefont {R.}~\bibnamefont {Thomale}},\ and\ \bibinfo {author} {\bibfnamefont {A.}~\bibnamefont {Szameit}},\ }\bibfield  {title} {\bibinfo {title} {Demonstration of a two-dimensional {{PT-symmetric}} crystal},\ }\href {https://doi.org/10.1038/s41467-018-08104-x} {\bibfield  {journal} {\bibinfo  {journal} {Nature Communications}\ }\textbf {\bibinfo {volume} {10}},\ \bibinfo {pages} {435} (\bibinfo {year} {2019})}\BibitemShut {NoStop}%
\bibitem [{\citenamefont {Miri}\ and\ \citenamefont {Al{\`u}}(2019)}]{miriExceptionalPointsOptics2019}%
  \BibitemOpen
  \bibfield  {author} {\bibinfo {author} {\bibfnamefont {M.-A.}\ \bibnamefont {Miri}}\ and\ \bibinfo {author} {\bibfnamefont {A.}~\bibnamefont {Al{\`u}}},\ }\bibfield  {title} {\bibinfo {title} {Exceptional points in optics and photonics},\ }\href {https://doi.org/10.1126/science.aar7709} {\bibfield  {journal} {\bibinfo  {journal} {Science}\ }\textbf {\bibinfo {volume} {363}},\ \bibinfo {pages} {eaar7709} (\bibinfo {year} {2019})}\BibitemShut {NoStop}%
\bibitem [{\citenamefont {{\"O}zdemir}\ \emph {et~al.}(2019)\citenamefont {{\"O}zdemir}, \citenamefont {Rotter}, \citenamefont {Nori},\ and\ \citenamefont {Yang}}]{ozdemirParityTimeSymmetry2019}%
  \BibitemOpen
  \bibfield  {author} {\bibinfo {author} {\bibfnamefont {{\c S}.~K.}\ \bibnamefont {{\"O}zdemir}}, \bibinfo {author} {\bibfnamefont {S.}~\bibnamefont {Rotter}}, \bibinfo {author} {\bibfnamefont {F.}~\bibnamefont {Nori}},\ and\ \bibinfo {author} {\bibfnamefont {L.}~\bibnamefont {Yang}},\ }\bibfield  {title} {\bibinfo {title} {Parity--time symmetry and exceptional points in photonics},\ }\href {https://doi.org/10.1038/s41563-019-0304-9} {\bibfield  {journal} {\bibinfo  {journal} {Nature Materials}\ }\textbf {\bibinfo {volume} {18}},\ \bibinfo {pages} {783} (\bibinfo {year} {2019})}\BibitemShut {NoStop}%
\bibitem [{\citenamefont {Khurgin}(2020)}]{khurginExceptionalPointsPolaritonic2020}%
  \BibitemOpen
  \bibfield  {author} {\bibinfo {author} {\bibfnamefont {J.~B.}\ \bibnamefont {Khurgin}},\ }\bibfield  {title} {\bibinfo {title} {Exceptional points in polaritonic cavities and subthreshold {{Fabry}}--{{Perot}} lasers},\ }\href {https://doi.org/10.1364/OPTICA.397378} {\bibfield  {journal} {\bibinfo  {journal} {Optica}\ }\textbf {\bibinfo {volume} {7}},\ \bibinfo {pages} {1015} (\bibinfo {year} {2020})}\BibitemShut {NoStop}%
\bibitem [{\citenamefont {McCreary}\ \emph {et~al.}(2017)\citenamefont {McCreary}, \citenamefont {Simpson}, \citenamefont {Wang}, \citenamefont {Rhodes}, \citenamefont {Fujisawa}, \citenamefont {Balicas}, \citenamefont {Dubey}, \citenamefont {Crespi}, \citenamefont {Terrones},\ and\ \citenamefont {Hight~Walker}}]{mccrearyIntricateResonantRaman2017}%
  \BibitemOpen
  \bibfield  {author} {\bibinfo {author} {\bibfnamefont {A.}~\bibnamefont {McCreary}}, \bibinfo {author} {\bibfnamefont {J.~R.}\ \bibnamefont {Simpson}}, \bibinfo {author} {\bibfnamefont {Y.}~\bibnamefont {Wang}}, \bibinfo {author} {\bibfnamefont {D.}~\bibnamefont {Rhodes}}, \bibinfo {author} {\bibfnamefont {K.}~\bibnamefont {Fujisawa}}, \bibinfo {author} {\bibfnamefont {L.}~\bibnamefont {Balicas}}, \bibinfo {author} {\bibfnamefont {M.}~\bibnamefont {Dubey}}, \bibinfo {author} {\bibfnamefont {V.~H.}\ \bibnamefont {Crespi}}, \bibinfo {author} {\bibfnamefont {M.}~\bibnamefont {Terrones}},\ and\ \bibinfo {author} {\bibfnamefont {A.~R.}\ \bibnamefont {Hight~Walker}},\ }\bibfield  {title} {\bibinfo {title} {Intricate {{Resonant Raman Response}} in {{Anisotropic ReS}} {\textsubscript{2}}},\ }\href {https://doi.org/10.1021/acs.nanolett.7b01463} {\bibfield  {journal} {\bibinfo  {journal} {Nano Letters}\ }\textbf {\bibinfo {volume} {17}},\ \bibinfo {pages} {5897} (\bibinfo {year} {2017})}\BibitemShut {NoStop}%
\bibitem [{\citenamefont {Pradhan}\ \emph {et~al.}(2015)\citenamefont {Pradhan}, \citenamefont {McCreary}, \citenamefont {Rhodes}, \citenamefont {Lu}, \citenamefont {Feng}, \citenamefont {Manousakis}, \citenamefont {Smirnov}, \citenamefont {Namburu}, \citenamefont {Dubey}, \citenamefont {Hight~Walker}, \citenamefont {Terrones}, \citenamefont {Terrones}, \citenamefont {Dobrosavljevic},\ and\ \citenamefont {Balicas}}]{pradhanMetalInsulatorQuantumPhase2015}%
  \BibitemOpen
  \bibfield  {author} {\bibinfo {author} {\bibfnamefont {N.~R.}\ \bibnamefont {Pradhan}}, \bibinfo {author} {\bibfnamefont {A.}~\bibnamefont {McCreary}}, \bibinfo {author} {\bibfnamefont {D.}~\bibnamefont {Rhodes}}, \bibinfo {author} {\bibfnamefont {Z.}~\bibnamefont {Lu}}, \bibinfo {author} {\bibfnamefont {S.}~\bibnamefont {Feng}}, \bibinfo {author} {\bibfnamefont {E.}~\bibnamefont {Manousakis}}, \bibinfo {author} {\bibfnamefont {D.}~\bibnamefont {Smirnov}}, \bibinfo {author} {\bibfnamefont {R.}~\bibnamefont {Namburu}}, \bibinfo {author} {\bibfnamefont {M.}~\bibnamefont {Dubey}}, \bibinfo {author} {\bibfnamefont {A.~R.}\ \bibnamefont {Hight~Walker}}, \bibinfo {author} {\bibfnamefont {H.}~\bibnamefont {Terrones}}, \bibinfo {author} {\bibfnamefont {M.}~\bibnamefont {Terrones}}, \bibinfo {author} {\bibfnamefont {V.}~\bibnamefont {Dobrosavljevic}},\ and\ \bibinfo {author} {\bibfnamefont {L.}~\bibnamefont {Balicas}},\ }\bibfield  {title} {\bibinfo {title} {Metal to {{Insulator Quantum-Phase Transition}} in
  {{Few-Layered ReS2}}},\ }\href {https://doi.org/10.1021/acs.nanolett.5b04100} {\bibfield  {journal} {\bibinfo  {journal} {Nano Letters}\ }\textbf {\bibinfo {volume} {15}},\ \bibinfo {pages} {8377} (\bibinfo {year} {2015})}\BibitemShut {NoStop}%
\bibitem [{\citenamefont {Feng}\ \emph {et~al.}(2015)\citenamefont {Feng}, \citenamefont {Zhou}, \citenamefont {Wang}, \citenamefont {Zhou}, \citenamefont {Liu}, \citenamefont {Fu}, \citenamefont {Ni}, \citenamefont {Wu}, \citenamefont {Yuan}, \citenamefont {Miao}, \citenamefont {Wang}, \citenamefont {Wan},\ and\ \citenamefont {Xing}}]{fengRamanVibrationalSpectra2015}%
  \BibitemOpen
  \bibfield  {author} {\bibinfo {author} {\bibfnamefont {Y.}~\bibnamefont {Feng}}, \bibinfo {author} {\bibfnamefont {W.}~\bibnamefont {Zhou}}, \bibinfo {author} {\bibfnamefont {Y.}~\bibnamefont {Wang}}, \bibinfo {author} {\bibfnamefont {J.}~\bibnamefont {Zhou}}, \bibinfo {author} {\bibfnamefont {E.}~\bibnamefont {Liu}}, \bibinfo {author} {\bibfnamefont {Y.}~\bibnamefont {Fu}}, \bibinfo {author} {\bibfnamefont {Z.}~\bibnamefont {Ni}}, \bibinfo {author} {\bibfnamefont {X.}~\bibnamefont {Wu}}, \bibinfo {author} {\bibfnamefont {H.}~\bibnamefont {Yuan}}, \bibinfo {author} {\bibfnamefont {F.}~\bibnamefont {Miao}}, \bibinfo {author} {\bibfnamefont {B.}~\bibnamefont {Wang}}, \bibinfo {author} {\bibfnamefont {X.}~\bibnamefont {Wan}},\ and\ \bibinfo {author} {\bibfnamefont {D.}~\bibnamefont {Xing}},\ }\bibfield  {title} {\bibinfo {title} {Raman vibrational spectra of bulk to monolayer
  {{ReS}}{\textsubscript{2}}} with lower symmetry},\ \href {https://doi.org/10.1103/PhysRevB.92.054110} {\bibfield  {journal} {\bibinfo  {journal} {Physical Review B}\ }\textbf {\bibinfo {volume} {92}},\ \bibinfo {pages} {054110} (\bibinfo {year} {2015})}\BibitemShut {NoStop}%
\bibitem [{\citenamefont {Jadczak}\ \emph {et~al.}(2019)\citenamefont {Jadczak}, \citenamefont {{Kutrowska-Girzycka}}, \citenamefont {Smole{\'n}ski}, \citenamefont {Kossacki}, \citenamefont {Huang},\ and\ \citenamefont {Bryja}}]{jadczakExcitonBindingEnergy2019}%
  \BibitemOpen
  \bibfield  {author} {\bibinfo {author} {\bibfnamefont {J.}~\bibnamefont {Jadczak}}, \bibinfo {author} {\bibfnamefont {J.}~\bibnamefont {{Kutrowska-Girzycka}}}, \bibinfo {author} {\bibfnamefont {T.}~\bibnamefont {Smole{\'n}ski}}, \bibinfo {author} {\bibfnamefont {P.}~\bibnamefont {Kossacki}}, \bibinfo {author} {\bibfnamefont {Y.~S.}\ \bibnamefont {Huang}},\ and\ \bibinfo {author} {\bibfnamefont {L.}~\bibnamefont {Bryja}},\ }\bibfield  {title} {\bibinfo {title} {Exciton binding energy and hydrogenic {{Rydberg}} series in layered {{ReS2}}},\ }\href {https://doi.org/10.1038/s41598-018-37655-8} {\bibfield  {journal} {\bibinfo  {journal} {Scientific Reports}\ }\textbf {\bibinfo {volume} {9}},\ \bibinfo {pages} {1578} (\bibinfo {year} {2019})}\BibitemShut {NoStop}%
\bibitem [{\citenamefont {Fainstein}\ \emph {et~al.}(1995)\citenamefont {Fainstein}, \citenamefont {Jusserand},\ and\ \citenamefont {{Thierry-Mieg}}}]{fainsteinRamanScatteringEnhancement1995}%
  \BibitemOpen
  \bibfield  {author} {\bibinfo {author} {\bibfnamefont {A.}~\bibnamefont {Fainstein}}, \bibinfo {author} {\bibfnamefont {B.}~\bibnamefont {Jusserand}},\ and\ \bibinfo {author} {\bibfnamefont {V.}~\bibnamefont {{Thierry-Mieg}}},\ }\bibfield  {title} {\bibinfo {title} {Raman {{Scattering Enhancement}} by {{Optical Confinement}} in a {{Semiconductor Planar Microcavity}}},\ }\href {https://doi.org/10.1103/PhysRevLett.75.3764} {\bibfield  {journal} {\bibinfo  {journal} {Physical Review Letters}\ }\textbf {\bibinfo {volume} {75}},\ \bibinfo {pages} {3764} (\bibinfo {year} {1995})}\BibitemShut {NoStop}%
\bibitem [{\citenamefont {Fainstein}\ \emph {et~al.}(1997)\citenamefont {Fainstein}, \citenamefont {Jusserand},\ and\ \citenamefont {{Thierry-Mieg}}}]{fainsteinCavityPolaritonMediatedResonant1997}%
  \BibitemOpen
  \bibfield  {author} {\bibinfo {author} {\bibfnamefont {A.}~\bibnamefont {Fainstein}}, \bibinfo {author} {\bibfnamefont {B.}~\bibnamefont {Jusserand}},\ and\ \bibinfo {author} {\bibfnamefont {V.}~\bibnamefont {{Thierry-Mieg}}},\ }\bibfield  {title} {\bibinfo {title} {Cavity-{{Polariton Mediated Resonant Raman Scattering}}},\ }\href {https://doi.org/10.1103/PhysRevLett.78.1576} {\bibfield  {journal} {\bibinfo  {journal} {Physical Review Letters}\ }\textbf {\bibinfo {volume} {78}},\ \bibinfo {pages} {1576} (\bibinfo {year} {1997})}\BibitemShut {NoStop}%
\bibitem [{See()}]{SeeSupplementalMaterial}%
  \BibitemOpen
  {}\bibinfo {note} {See Supplemental Material provided as ancillary file at  https://arxiv.org/abs/2305.17475 for additional data and theory.}\BibitemShut {Stop}%
\bibitem [{\citenamefont {Pattanayak}\ \emph {et~al.}(2022)\citenamefont {Pattanayak}, \citenamefont {Das}, \citenamefont {Dhara}, \citenamefont {Chakrabarty}, \citenamefont {Paul}, \citenamefont {Gurnani}, \citenamefont {Brundavanam},\ and\ \citenamefont {Dhara}}]{pattanayakSteadyStateApproachStudying2022}%
  \BibitemOpen
  \bibfield  {author} {\bibinfo {author} {\bibfnamefont {A.~K.}\ \bibnamefont {Pattanayak}}, \bibinfo {author} {\bibfnamefont {P.}~\bibnamefont {Das}}, \bibinfo {author} {\bibfnamefont {A.}~\bibnamefont {Dhara}}, \bibinfo {author} {\bibfnamefont {D.}~\bibnamefont {Chakrabarty}}, \bibinfo {author} {\bibfnamefont {S.}~\bibnamefont {Paul}}, \bibinfo {author} {\bibfnamefont {K.}~\bibnamefont {Gurnani}}, \bibinfo {author} {\bibfnamefont {M.~M.}\ \bibnamefont {Brundavanam}},\ and\ \bibinfo {author} {\bibfnamefont {S.}~\bibnamefont {Dhara}},\ }\bibfield  {title} {\bibinfo {title} {A {{Steady-State Approach}} for {{Studying Valley Relaxation Using}} an {{Optical Vortex Beam}}},\ }\bibfield  {journal} {\bibinfo  {journal} {Nano Letters}\ }\href {https://doi.org/10.1021/acs.nanolett.2c00824} {10.1021/acs.nanolett.2c00824} (\bibinfo {year} {2022})\BibitemShut {NoStop}%
\bibitem [{\citenamefont {Zhou}\ \emph {et~al.}(2020)\citenamefont {Zhou}, \citenamefont {Maity}, \citenamefont {Rai}, \citenamefont {Juneja}, \citenamefont {Meng}, \citenamefont {Roy}, \citenamefont {Zhang}, \citenamefont {Xu}, \citenamefont {Lin}, \citenamefont {Banerjee}, \citenamefont {Singh},\ and\ \citenamefont {Wang}}]{zhouStackingOrderDrivenOpticalProperties2020}%
  \BibitemOpen
  \bibfield  {author} {\bibinfo {author} {\bibfnamefont {Y.}~\bibnamefont {Zhou}}, \bibinfo {author} {\bibfnamefont {N.}~\bibnamefont {Maity}}, \bibinfo {author} {\bibfnamefont {A.}~\bibnamefont {Rai}}, \bibinfo {author} {\bibfnamefont {R.}~\bibnamefont {Juneja}}, \bibinfo {author} {\bibfnamefont {X.}~\bibnamefont {Meng}}, \bibinfo {author} {\bibfnamefont {A.}~\bibnamefont {Roy}}, \bibinfo {author} {\bibfnamefont {Y.}~\bibnamefont {Zhang}}, \bibinfo {author} {\bibfnamefont {X.}~\bibnamefont {Xu}}, \bibinfo {author} {\bibfnamefont {J.-F.}\ \bibnamefont {Lin}}, \bibinfo {author} {\bibfnamefont {S.~K.}\ \bibnamefont {Banerjee}}, \bibinfo {author} {\bibfnamefont {A.~K.}\ \bibnamefont {Singh}},\ and\ \bibinfo {author} {\bibfnamefont {Y.}~\bibnamefont {Wang}},\ }\bibfield  {title} {\bibinfo {title} {Stacking-{{Order}}-{{Driven Optical Properties}} and {{Carrier Dynamics}} in {{ReS}} {\textsubscript{2}}},\ }\href {https://doi.org/10.1002/adma.201908311} {\bibfield  {journal} {\bibinfo  {journal} {Advanced
  Materials}\ }\textbf {\bibinfo {volume} {32}},\ \bibinfo {pages} {1908311} (\bibinfo {year} {2020})}\BibitemShut {NoStop}%
\bibitem [{\citenamefont {McKeever}\ \emph {et~al.}(2003)\citenamefont {McKeever}, \citenamefont {Boca}, \citenamefont {Boozer}, \citenamefont {Buck},\ and\ \citenamefont {Kimble}}]{mckeeverExperimentalRealizationOneatom2003}%
  \BibitemOpen
  \bibfield  {author} {\bibinfo {author} {\bibfnamefont {J.}~\bibnamefont {McKeever}}, \bibinfo {author} {\bibfnamefont {A.}~\bibnamefont {Boca}}, \bibinfo {author} {\bibfnamefont {A.~D.}\ \bibnamefont {Boozer}}, \bibinfo {author} {\bibfnamefont {J.~R.}\ \bibnamefont {Buck}},\ and\ \bibinfo {author} {\bibfnamefont {H.~J.}\ \bibnamefont {Kimble}},\ }\bibfield  {title} {\bibinfo {title} {Experimental realization of a one-atom laser in the regime of strong coupling},\ }\href {https://doi.org/10.1038/nature01974} {\bibfield  {journal} {\bibinfo  {journal} {Nature}\ }\textbf {\bibinfo {volume} {425}},\ \bibinfo {pages} {268} (\bibinfo {year} {2003})}\BibitemShut {NoStop}%
\bibitem [{\citenamefont {Yariv}(1989)}]{yarivQuantumElectronics1989}%
  \BibitemOpen
  \bibfield  {author} {\bibinfo {author} {\bibfnamefont {A.}~\bibnamefont {Yariv}},\ }\href@noop {} {\emph {\bibinfo {title} {Quantum Electronics}}},\ \bibinfo {edition} {3rd}\ ed.\ (\bibinfo  {publisher} {Wiley},\ \bibinfo {address} {New York},\ \bibinfo {year} {1989})\BibitemShut {NoStop}%
\bibitem [{\citenamefont {Wu}\ \emph {et~al.}(1999)\citenamefont {Wu}, \citenamefont {Yang},\ and\ \citenamefont {Leung}}]{wuTheoryMicrocavityenhancedRaman1999}%
  \BibitemOpen
  \bibfield  {author} {\bibinfo {author} {\bibfnamefont {Y.}~\bibnamefont {Wu}}, \bibinfo {author} {\bibfnamefont {X.}~\bibnamefont {Yang}},\ and\ \bibinfo {author} {\bibfnamefont {P.~T.}\ \bibnamefont {Leung}},\ }\bibfield  {title} {\bibinfo {title} {Theory of microcavity-enhanced {{Raman}} gain},\ }\href {https://doi.org/10.1364/OL.24.000345} {\bibfield  {journal} {\bibinfo  {journal} {Optics Letters}\ }\textbf {\bibinfo {volume} {24}},\ \bibinfo {pages} {345} (\bibinfo {year} {1999})}\BibitemShut {NoStop}%
\bibitem [{\citenamefont {Yamamoto}\ and\ \citenamefont {Slusher}(1993)}]{yamamotoOpticalProcessesMicrocavities1993}%
  \BibitemOpen
  \bibfield  {author} {\bibinfo {author} {\bibfnamefont {Y.}~\bibnamefont {Yamamoto}}\ and\ \bibinfo {author} {\bibfnamefont {R.~E.}\ \bibnamefont {Slusher}},\ }\bibfield  {title} {\bibinfo {title} {Optical {{Processes}} in {{Microcavities}}},\ }\href {https://doi.org/10.1063/1.881356} {\bibfield  {journal} {\bibinfo  {journal} {Physics Today}\ }\textbf {\bibinfo {volume} {46}},\ \bibinfo {pages} {66} (\bibinfo {year} {1993})}\BibitemShut {NoStop}%
\bibitem [{\citenamefont {Khajavikhan}\ \emph {et~al.}(2012)\citenamefont {Khajavikhan}, \citenamefont {Simic}, \citenamefont {Katz}, \citenamefont {Lee}, \citenamefont {Slutsky}, \citenamefont {Mizrahi}, \citenamefont {Lomakin},\ and\ \citenamefont {Fainman}}]{khajavikhanThresholdlessNanoscaleCoaxial2012}%
  \BibitemOpen
  \bibfield  {author} {\bibinfo {author} {\bibfnamefont {M.}~\bibnamefont {Khajavikhan}}, \bibinfo {author} {\bibfnamefont {A.}~\bibnamefont {Simic}}, \bibinfo {author} {\bibfnamefont {M.}~\bibnamefont {Katz}}, \bibinfo {author} {\bibfnamefont {J.~H.}\ \bibnamefont {Lee}}, \bibinfo {author} {\bibfnamefont {B.}~\bibnamefont {Slutsky}}, \bibinfo {author} {\bibfnamefont {A.}~\bibnamefont {Mizrahi}}, \bibinfo {author} {\bibfnamefont {V.}~\bibnamefont {Lomakin}},\ and\ \bibinfo {author} {\bibfnamefont {Y.}~\bibnamefont {Fainman}},\ }\bibfield  {title} {\bibinfo {title} {Thresholdless nanoscale coaxial lasers},\ }\href {https://doi.org/10.1038/nature10840} {\bibfield  {journal} {\bibinfo  {journal} {Nature}\ }\textbf {\bibinfo {volume} {482}},\ \bibinfo {pages} {204} (\bibinfo {year} {2012})}\BibitemShut {NoStop}%
\bibitem [{\citenamefont {Prieto}\ \emph {et~al.}(2015)\citenamefont {Prieto}, \citenamefont {Llorens}, \citenamefont {{Mu{\~n}oz-Cam{\'u}{\~n}ez}}, \citenamefont {Taboada}, \citenamefont {{Canet-Ferrer}}, \citenamefont {Ripalda}, \citenamefont {Robles}, \citenamefont {{Mu{\~n}oz-Matutano}}, \citenamefont {{Mart{\'i}nez-Pastor}},\ and\ \citenamefont {Postigo}}]{prietoThresholdlessLaserOperation2015}%
  \BibitemOpen
  \bibfield  {author} {\bibinfo {author} {\bibfnamefont {I.}~\bibnamefont {Prieto}}, \bibinfo {author} {\bibfnamefont {J.~M.}\ \bibnamefont {Llorens}}, \bibinfo {author} {\bibfnamefont {L.~E.}\ \bibnamefont {{Mu{\~n}oz-Cam{\'u}{\~n}ez}}}, \bibinfo {author} {\bibfnamefont {A.~G.}\ \bibnamefont {Taboada}}, \bibinfo {author} {\bibfnamefont {J.}~\bibnamefont {{Canet-Ferrer}}}, \bibinfo {author} {\bibfnamefont {J.~M.}\ \bibnamefont {Ripalda}}, \bibinfo {author} {\bibfnamefont {C.}~\bibnamefont {Robles}}, \bibinfo {author} {\bibfnamefont {G.}~\bibnamefont {{Mu{\~n}oz-Matutano}}}, \bibinfo {author} {\bibfnamefont {J.~P.}\ \bibnamefont {{Mart{\'i}nez-Pastor}}},\ and\ \bibinfo {author} {\bibfnamefont {P.~A.}\ \bibnamefont {Postigo}},\ }\bibfield  {title} {\bibinfo {title} {Near thresholdless laser operation at room temperature},\ }\href {https://doi.org/10.1364/OPTICA.2.000066} {\bibfield  {journal} {\bibinfo  {journal} {Optica}\ }\textbf {\bibinfo {volume} {2}},\ \bibinfo {pages} {66} (\bibinfo {year}
  {2015})}\BibitemShut {NoStop}%
\bibitem [{\citenamefont {Wu}\ \emph {et~al.}(2017)\citenamefont {Wu}, \citenamefont {Park}, \citenamefont {Lim},\ and\ \citenamefont {Klimov}}]{wuZerothresholdOpticalGain2017}%
  \BibitemOpen
  \bibfield  {author} {\bibinfo {author} {\bibfnamefont {K.}~\bibnamefont {Wu}}, \bibinfo {author} {\bibfnamefont {Y.-S.}\ \bibnamefont {Park}}, \bibinfo {author} {\bibfnamefont {J.}~\bibnamefont {Lim}},\ and\ \bibinfo {author} {\bibfnamefont {V.~I.}\ \bibnamefont {Klimov}},\ }\bibfield  {title} {\bibinfo {title} {Towards zero-threshold optical gain using charged semiconductor quantum dots},\ }\href {https://doi.org/10.1038/nnano.2017.189} {\bibfield  {journal} {\bibinfo  {journal} {Nature Nanotechnology}\ }\textbf {\bibinfo {volume} {12}},\ \bibinfo {pages} {1140} (\bibinfo {year} {2017})}\BibitemShut {NoStop}%
\bibitem [{\citenamefont {Jagsch}\ \emph {et~al.}(2018)\citenamefont {Jagsch}, \citenamefont {Trivi{\~n}o}, \citenamefont {Lohof}, \citenamefont {Callsen}, \citenamefont {Kalinowski}, \citenamefont {Rousseau}, \citenamefont {Barzel}, \citenamefont {Carlin}, \citenamefont {Jahnke}, \citenamefont {Butt{\'e}}, \citenamefont {Gies}, \citenamefont {Hoffmann}, \citenamefont {Grandjean},\ and\ \citenamefont {Reitzenstein}}]{jagschQuantumOpticalStudy2018}%
  \BibitemOpen
  \bibfield  {author} {\bibinfo {author} {\bibfnamefont {S.~T.}\ \bibnamefont {Jagsch}}, \bibinfo {author} {\bibfnamefont {N.~V.}\ \bibnamefont {Trivi{\~n}o}}, \bibinfo {author} {\bibfnamefont {F.}~\bibnamefont {Lohof}}, \bibinfo {author} {\bibfnamefont {G.}~\bibnamefont {Callsen}}, \bibinfo {author} {\bibfnamefont {S.}~\bibnamefont {Kalinowski}}, \bibinfo {author} {\bibfnamefont {I.~M.}\ \bibnamefont {Rousseau}}, \bibinfo {author} {\bibfnamefont {R.}~\bibnamefont {Barzel}}, \bibinfo {author} {\bibfnamefont {J.-F.}\ \bibnamefont {Carlin}}, \bibinfo {author} {\bibfnamefont {F.}~\bibnamefont {Jahnke}}, \bibinfo {author} {\bibfnamefont {R.}~\bibnamefont {Butt{\'e}}}, \bibinfo {author} {\bibfnamefont {C.}~\bibnamefont {Gies}}, \bibinfo {author} {\bibfnamefont {A.}~\bibnamefont {Hoffmann}}, \bibinfo {author} {\bibfnamefont {N.}~\bibnamefont {Grandjean}},\ and\ \bibinfo {author} {\bibfnamefont {S.}~\bibnamefont {Reitzenstein}},\ }\bibfield  {title} {\bibinfo {title} {A quantum optical study of thresholdless lasing
  features in high-{$\beta$} nitride nanobeam cavities},\ }\href {https://doi.org/10.1038/s41467-018-02999-2} {\bibfield  {journal} {\bibinfo  {journal} {Nature Communications}\ }\textbf {\bibinfo {volume} {9}},\ \bibinfo {pages} {564} (\bibinfo {year} {2018})}\BibitemShut {NoStop}%
\bibitem [{\citenamefont {Checoury}\ \emph {et~al.}(2010)\citenamefont {Checoury}, \citenamefont {Han}, \citenamefont {El~Kurdi},\ and\ \citenamefont {Boucaud}}]{checouryDeterministicMeasurementPurcell2010}%
  \BibitemOpen
  \bibfield  {author} {\bibinfo {author} {\bibfnamefont {X.}~\bibnamefont {Checoury}}, \bibinfo {author} {\bibfnamefont {Z.}~\bibnamefont {Han}}, \bibinfo {author} {\bibfnamefont {M.}~\bibnamefont {El~Kurdi}},\ and\ \bibinfo {author} {\bibfnamefont {P.}~\bibnamefont {Boucaud}},\ }\bibfield  {title} {\bibinfo {title} {Deterministic measurement of the {{Purcell}} factor in microcavities through {{Raman}} emission},\ }\href {https://doi.org/10.1103/PhysRevA.81.033832} {\bibfield  {journal} {\bibinfo  {journal} {Physical Review A}\ }\textbf {\bibinfo {volume} {81}},\ \bibinfo {pages} {033832} (\bibinfo {year} {2010})}\BibitemShut {NoStop}%
\bibitem [{\citenamefont {Petrak}\ \emph {et~al.}(2014)\citenamefont {Petrak}, \citenamefont {Djeu},\ and\ \citenamefont {Muller}}]{petrakPurcellenhancedRamanScattering2014}%
  \BibitemOpen
  \bibfield  {author} {\bibinfo {author} {\bibfnamefont {B.}~\bibnamefont {Petrak}}, \bibinfo {author} {\bibfnamefont {N.}~\bibnamefont {Djeu}},\ and\ \bibinfo {author} {\bibfnamefont {A.}~\bibnamefont {Muller}},\ }\bibfield  {title} {\bibinfo {title} {Purcell-enhanced {{Raman}} scattering from atmospheric gases in a high-finesse microcavity},\ }\href {https://doi.org/10.1103/PhysRevA.89.023811} {\bibfield  {journal} {\bibinfo  {journal} {Physical Review A}\ }\textbf {\bibinfo {volume} {89}},\ \bibinfo {pages} {023811} (\bibinfo {year} {2014})}\BibitemShut {NoStop}%
\bibitem [{\citenamefont {Kavokin}\ \emph {et~al.}(2017)\citenamefont {Kavokin}, \citenamefont {Baumberg}, \citenamefont {Malpuech},\ and\ \citenamefont {Laussy}}]{kavokinMicrocavities2017}%
  \BibitemOpen
  \bibfield  {author} {\bibinfo {author} {\bibfnamefont {A.~V.}\ \bibnamefont {Kavokin}}, \bibinfo {author} {\bibfnamefont {J.}~\bibnamefont {Baumberg}}, \bibinfo {author} {\bibfnamefont {G.}~\bibnamefont {Malpuech}},\ and\ \bibinfo {author} {\bibfnamefont {F.~P.}\ \bibnamefont {Laussy}},\ }\href@noop {} {\emph {\bibinfo {title} {{Microcavities}}}},\ \bibinfo {edition} {2nd ed.}\,\ \bibinfo {series} {{Series on semiconductor science and technology}}\ No.~\bibinfo {number} {21}\ (\bibinfo  {publisher} {Oxford University Press},\ \bibinfo {address} {Oxford},\ \bibinfo {year} {2017})\BibitemShut {NoStop}%
\end{thebibliography}
%

\end{document}